\documentclass[prb,nofootinbib,raggedfooter,preprintnumbers]{revtex4}
\usepackage{amsfonts,amsmath}

\begin      {document}
\title      {Exact replica treatment of non-Hermitean complex
                random matrices\footnote{
                Published in: \textsf{Frontiers in Field
                Theory}, edited by O. Kovras, Ch. 3, pp. 25 -- 54 (Nova Science Publishers Inc.,
                New York, 2004). ISBN: 1-59454-127-2.
                }
            }
\preprint   {{\it Dedicated to the memory of Professor Iya
                  Ipatova}}
\author     {Eugene Kanzieper\footnote{Electronic address: \textsf{eugene.kanzieper@weizmann.ac.il} }}
\affiliation{Department of Applied
Mathematics, School of Science\\ Holon Academic Institute of
             Technology, Holon 58102, Israel
            }
\date       {November 29, 2003}

\begin{abstract}
Recently discovered exact integrability of zero-dimensional
replica field theories [E. Kanzieper, Phys. Rev. Lett. {\bf 89},
250201 (2002)] is examined in the context of Ginibre Unitary
Ensemble of non-Hermitean random matrices (GinUE). In particular,
various nonperturbative fermionic replica partition functions for
this random matrix model are shown to belong to a positive, {\it
semi-infinite} Toda Lattice Hierarchy which, upon its Painlev\'e
reduction, yields exact expressions for the mean level density and
the density-density correlation function in both bulk of the
complex spectrum and near its edges. Comparison is made with an
approximate treatment of non-Hermitean disordered Hamiltonians
based on the `replica symmetry breaking' ansatz. A difference
between our replica approach and a framework exploiting the
replica limit of an {\it infinite} (supersymmetric) Toda Lattice
equation is also discussed.
\end{abstract}

\maketitle

\section{Introduction}
\label{Sec.Intro} \textbf{\emph{How replicas arise.}}---In physics
of disorder, all observables depend in highly nonlinear fashion on
a stochastic Hamiltonian hereby making calculation of their
ensemble averages very difficult. To determine the latter in an
interactionless system, one has to know spectral statistical
properties of a single particle Hamiltonian ${\cal H}$ contained
in the mean product of resolvents, $G(\varepsilon) = {\rm tr}
\left( \varepsilon - {\cal H} \right)^{-1}$. Each of the
resolvents can exactly be represented as a ratio of two integrals
running over an auxiliary vector field $\psi$ which may consist of
either commuting (bosonic) or anticommuting (fermionic) entries.
In the random matrix theory limit, when a system Hamiltonian is
modelled by an $N \times N$ random matrix ${\cal H}$ of certain
symmetries, the resolvent $G(\varepsilon)$ equals
\begin{eqnarray}
    \label{ratio}
    G(\varepsilon)
    &=&
    i \int {\cal D}{\bar \psi}  {\cal D} \psi \;
    {\bar \psi}_\ell \psi_\ell \, e^{
    -i S_{\cal H}[\varepsilon;\, {\bar \psi}, \psi]}
    \left(
    \int {\cal D}{\bar \psi}  {\cal D} \psi \;
    e^{
    -i S_{\cal H}[\varepsilon; \, {\bar \psi}, \psi]
    }
    \right)^{-1}
\end{eqnarray}
where $
    S_{\cal H}=
    {\bar \psi}_\ell
    \left( \varepsilon \delta_{\ell\, \ell^\prime}-{\cal H}_{\ell\, \ell^\prime} \right)
    \psi_{\ell^\prime}
    $, $\psi$ is an $N$--component vector $\psi^{\rm T} = (\psi_1,\cdots,
    \psi_N)$, ${\bar \psi}$ is its proper conjugate, and ${\rm Im\,} \varepsilon
\ne 0$. Summation over repeated Latin indices is assumed.

Although exact, this representation is a bit too inconvenient for
a nonperturbative averaging due to the awkward random denominator.
To get rid of it, Edwards and Anderson \cite{EA-1975} proposed a
replica method based on the identity
\begin{eqnarray}
    \label{log-Z}
    \ln \, Z
    =
    \lim_{n \rightarrow \pm 0}
    \frac{Z^{n} - 1}{n}.
\end{eqnarray}
Upon assigning to $Z$ a meaning of a quantum partition function
\begin{eqnarray}
    \label{partition}
    Z(\varepsilon) = i^N
    \int {\cal D}{\bar \psi}  {\cal D} \psi \;
    e^{
    - i S_{\cal H}[\varepsilon; \, {\bar \psi}, \psi]
    },
\end{eqnarray}
the average resolvent $\langle G(\varepsilon) \rangle$ can be
determined through the limiting procedure
\begin{eqnarray}
    \label{r-limit}
    \langle G(\varepsilon) \rangle = \lim_{n \rightarrow 0} \frac{1}{n} \,
    \frac{\partial}{\partial \varepsilon} \, \langle
    Z^n(\varepsilon)\rangle
\end{eqnarray}
involving the average of the partition function
\begin{equation}
    \label{replica}
    Z^n(\varepsilon)= i^{nN}
    \int \prod_{\alpha=1}^{|n|}
    {\cal D}{\bar \psi}^{(\alpha)}  {\cal D} \psi^{(\alpha)} \;
    e^{
    - i S_{\cal H}
    [\varepsilon;
    \, {\bar \psi}^{(\alpha)},
    \psi^{(\alpha)}
    ]
    }
\end{equation}
describing $n$ identical noninteracting copies, or replicas, of
the initial disordered system (\ref{partition}). The nature of
replicated fields $\psi^{(\alpha)}$ in (\ref{replica}) is
determined by the sign of $n$ in (\ref{log-Z}); they are bosonic
for $n < 0$ and are fermionic otherwise. Clearly, the procedure
(\ref{r-limit}) assumes mutual commutativity of the replica limit
$n \rightarrow 0$, an ensemble averaging $\langle \cdots \rangle$,
and the differentiation operation $\partial/\partial \varepsilon$.

Contrary to (\ref{ratio}), the representations (\ref{r-limit}) and
(\ref{replica}) contain no random denominator hereby making a
nonperturbative in disorder calculation of the resolvent
$G(\varepsilon)$ viable. Depending on the origin of auxiliary
fields $\psi^{(\alpha)}$ in (\ref{replica}), such a disorder
averaging supplemented by identification of physically relevant
low lying modes of the theory, would eventually result in
effective replica field theory\cite{Remark-example} (also called a
nonlinear replica $\sigma$ model) defined on either a noncompact
\cite{W-1979,SW-1980} or a compact \cite{ELK-1980} manifold.

\textbf{\emph{Why the replica limit is problematic.}}---Seemingly
innocent at first glance, the above field theoretic construction
appears to be counterintuitive and rising fundamental mathematical
questions\cite{P-2002}. Indeed, due to a particular integration
measure which makes no sense for $n$ other than integers, the
average of (\ref{replica}) cannot directly be used to implement
the replica limit (\ref{r-limit}) determined by the behaviour of
$Z_n(\varepsilon)=\langle Z^n(\varepsilon) \rangle$ in a close
vicinity of $n=0$. To circumvent this difficulty (which
reflects\cite{K-2003} a true, {\it continuous geometry}\,
\cite{P-2002} of replica field theories), one may at first
evaluate the average replica partition function $Z_n(\varepsilon)$
for all $n \in {\mathbb Z^+}$ (or ${\mathbb Z^-}$), and then
analytically continue the result into a vicinity of $n=0$ in order
to make taking the replica limit (\ref{r-limit}) well defined and
safe. This route, however, is full of pitfalls.

In the context of mesoscopic physics, the subtleties involved in
carrying out the replica limit were discussed at length in Refs.
\cite{VZ-1985,KM-1999a,KM-1999b,YL-1999,Z-1999,K-2001}. All these
studies have debated {\it whether or not the nonperturbative
\cite{Remark-perturbative} sector of replica field theories is
reliable}. This issue is of conceptual importance yet is not pure
academic because the replica field theories are among a very few
means available to address problems involving both disorder and
interactions, about which the famous Efetov's supersymmetry approach
(SUSY) -- a prime tool in studying noninteracting disordered systems
for the last two decades -- has nothing to say beyond a
weak-interaction limit \cite{Remark-Keldysh}.

\textbf{\emph{Approximate treatment of replicas.}}---In the early
study \cite{VZ-1985} by Verbaarschot and Zirnbauer, a
nonperturbative sector of nonlinear replica $\sigma$ models was
thoroughly examined in the context of the Random Matrix Theory
\cite{M-1991} (RMT). Having mapped the problem of eigenvalue
correlations in the Gaussian Unitary Ensemble (GUE) of large
random matrices onto both bosonic and fermionic replica field
theories, these authors had found that the two formulations of
nonlinear replica $\sigma$ model supplied different results for
the density-density correlation function, both apparently
differing from the correct one firmly established by other
methods\cite{M-1991,E-book}. These findings led the authors to
conclude that the replica method is `mathematically ill founded'.
The failure of the replica method to correctly account for all
nonperturbative contributions to a physical observable was
attributed to a {\it nonuniqueness} of the analytic continuation
of replica partition functions in the replica parameter $n$ away
from (either negative or positive) integers. This standpoint,
recently reiterated by Zirnbauer \cite{Z-1999}, has formed a
prevailing opinion in the literature that the replica method may
at best be considered as a perturbative tool not being able to
reproduce truly nonperturbative results accessible by alternative
SUSY technique.

The paper that challenged the opinion about inner deficiency of
replica field theories and triggered their further reassessment
was that of Kamenev and M\'ezard \cite{KM-1999a}. Based on ideas
of replica symmetry breaking originally devised in the theory of
spin glasses \cite{MPV-1987}, these authors came up with a
procedure that eventually produced nonperturbative results for the
GUE density-density correlation function out of fermionic
replicas, albeit in an asymptotic region describing evolution of a
quantum system at times not exceeding the Heisenberg time.
(Subsequently, this approach was applied to a number of problems
such as the energy level statistics in disordered metallic grains
beyond \cite{KM-1999b} the RMT limit, spatial correlations in
Calogero-Sutherland models \cite{GK-2001,NGK-2002}, a microscopic
spectral density of the Euclidean QCD Dirac operator
\cite{DV-2001}, and energy level fluctuations in Ginibre ensembles
of non-Hermitean random matrices \cite{NK-2002}).

Briefly summarised (for a detailed exposition the reader is
referred to original publications \cite{KM-1999a,KM-1999b} as well
as to a critical analysis \cite{Z-1999} by Zirnbauer), the
framework \cite{KM-1999a} rests on an {\it approximate} saddle
point evaluation of replica partition functions represented in
terms of $|n|$--fold integrals containing a large parameter. In
doing so, nontrivial saddle point configurations with so-called
broken replica symmetry have to be taken into account in order to
reproduce nonperturbative results. While leading to asymptotically
correct expressions for spectral fluctuations in the Gaussian
ensembles possessing unitary, orthogonal and symplectic Dyson's
symmetries, the procedure\cite{KM-1999a} cannot be considered as
mathematically satisfactory because it involves a nonexisting
\cite{KM-1999a,Z-1999} analytic continuation of replica partition
functions to a vicinity of $n = 0$, the domain which is crucially
important for implementing the replica limit.

\textbf{\emph{Towards exact integrability of replica
$\boldsymbol{\sigma}$ models.}}---Analysis of Refs.
\cite{VZ-1985,KM-1999a,KM-1999b,YL-1999,Z-1999} (see also Ref.
\cite{K-2001}) hints that {\it approximate} evaluation of replica
partition functions is the key point\cite{K-2002} to blame for
inconsistencies encountered in the procedure of analytic
continuation away from $n$ integers. In such a situation, leaning
towards {\it exact} calculational schemes in replica field
theories is a natural move.

A step in this direction was taken in the recent paper
\cite{K-2002}, where partition functions for a number of fermionic
replica $\sigma$ models were shown to belong to a {\it positive,
semi-infinite} Toda Lattice Hierarchy extensively studied in the
theory of nonlinear integrable systems \cite{T-2000}. In
conjunction with the $\tau$-function theory \cite{O-1,O-2,O-3,O-4}
of the six Painlev\'e transcendents (which are yet another
fundamental object in the theory of integrable hierarchies), this
observation led to {\it exact} evaluation of replica partition
functions for a number of random matrix ensembles in terms of
Painlev\'e transcendents. Resulting nonperturbative Painlev\'e
representations (which implicitly encode {\it all} hierarchical
inter-relations between partition functions with various replica
indices) were used to build a continuation of $Z_n$'s away from $n
\in {\mathbb Z^+}$. While not addressing the important issue of
{\it uniqueness} of such an analytic continuation, the route of
Ref. \cite{K-2002} has yielded -- for the first time -- exact
nonperturbative results for random matrix spectral statistics out
of fermionic replicas.

More recently, Splittorff and Verbaarschot
\cite{SV-2003,SV-2003new} have suggested that such nonperturbative
results could directly be obtained from the replica limit of an
{\it infinite} Toda Lattice equation without Painlev\'e reduction
whatsoever. In fact, the approach \cite{SV-2003,SV-2003new} rests
on the observation that, if properly normalised, the fermionic and
the bosonic replica partition functions of a zero-dimensional
interactionless system form a {\it single, infinite} (that is,
supersymmetric) Toda Lattice Hierarchy belonging to its either
positive (fermionic) or negative (bosonic) branch. While greatly
simplifying calculations of spectral correlation functions through
a remarkable bosonic-fermionic factorisation\cite{SV-2003new}, the
framework developed by Splittorff and Verbaarschot is, to a large
extent, supersymmetric in nature as it explicitly
injects\cite{New-Remark} a missing bosonic (or fermionic)
information to otherwise fermionic (or bosonic) like treatment.

In the present paper, a detailed account is offered of a
nonperturbative approach \cite{K-2002} to zero dimensional
fermionic replica field theories which is based on exact
Painlev\'e representation of replica partition functions with the
emphasis strongly placed on technical details. Specifically, we
focus on the Ginibre ensemble \cite{G-1965} of complex random
matrices with no further symmetries. This particular random matrix
model is of special interest in the light of recent findings that
associate statistical models of normal random matrices with
integrable structures of conformal maps and interface dynamics at
both classical \cite{MWZ-2000} and quantum scales
\cite{ABWZ-2002}. (The reader is referred to Ref. \cite{Z-2002}
for an introductory exposition of these recent developments and to
Ref. \cite{FS-2003} for a review of other physical applications
and extended bibliography).

The paper is organised as follows. In Sec. II, we collect the
basic definitions and present the major results regarding the
Ginibre Unitary Ensemble of random matrices. In Sec. III, a
fermionic replica field theory approach to non-Hermitean complex
random matrices is outlined and integrability of the field theory
is established. The integrability which manifests itself in
emergence of a positive, semi-infinite Toda Lattice equation for
replica partition functions and also results in exact
representations of replica partition functions in terms of
Painlev\'e transcendents, eventually culminates in reproducing
exact fluctuation formulas for Ginibre complex random matrices.
Finally, in Sec. IV, we make a comparison of our exact approach
with the approximate treatment\cite{NK-2002} of non-Hermitean
disordered Hamiltonians based on the `replica symmetry breaking'
ansatz; we also comment on differences between our approach to
replicas and a framework\cite{SV-2003,SV-2003new} exploiting a
replica limit of the infinite Toda Lattice equation. Most lengthy
technical details are collected in the two appendices, A and B.

\section{Ginibre unitary ensemble of non-Hermitean random matrices:
Definitions and basic results} \label{Sec.Mehta}

\textbf{\emph{Preliminaries.}}---Statistical ensemble of generic
$N\times N$ complex random matrices ${\cal H} \in {\mathbb
C}^{N\times N}$ whose entries are independently distributed in
accordance with the Gaussian law \cite{Remark-NC}
\begin{eqnarray}
    \label{P-of-H}
    P_N({\cal H}) =\pi^{-N^2} \exp\left( - {\rm tr}\; {\cal H} {\cal H}^\dagger
    \right)
\end{eqnarray}
has first been introduced in the pioneering work by Ginibre
\cite{G-1965}. Throughout the paper, such random Hamiltonians will
be denoted as ${\cal H} \in {\rm GinUE}_N$.

The joint probability distribution function $P_N (z_1,\cdots,z_N)$
of $N$ eigenvalues of the matrix ${\cal H}$ is of particular
interest. Having Schur-decomposed the ${\cal H}$ as ${\cal H} =
U^\dagger ({\cal Z}+{\cal R})\,U$ where $U$ is a unitary matrix
which is unique up to the phase of each column, ${\cal R}$ is a
strictly upper-triangular complex matrix, and ${\cal Z}$ is a
diagonal matrix ${\cal Z}={\rm diag}\, (z_1,\cdots, z_N)$
consisting of $N$ complex eigenvalues $\{z_\ell\} = \{x_\ell + i
y_\ell\}$ of the ${\cal H}$, Ginibre managed to derive the joint
probability distribution function of $\{z_\ell\}$ in the form
\begin{equation}
\label{jpdf}
    P_N (z_1,\cdots,z_N) = C(N) \prod_{\ell_1<\ell_2=1}^N
    |z_{\ell_1} - z_{\ell_2}|^2 \prod_{\ell=1}^N
    w^2(z_\ell,\bar{z}_\ell)
\end{equation}
where the weight function $w^2(z,\bar{z})$ equals $w^2(z,\bar{z})
= \exp(-z\bar{z})$. Given the integration measure $d^2 Z_\ell =
dx_\ell dy_\ell$, the inverse normalisation constant in
(\ref{jpdf}) is determined to be $C^{-1}(N)= \pi^N \Gamma(N+1)$.

Of primary interest is the $p$-point correlation function
\begin{equation}
    \label{p-point}
    R_p(z_1,\ldots,z_p;N) = \frac{N!}{(N-p)!} \int
    \prod_{\ell=p+1}^N
    d^2 Z_\ell \, P_N(z_1,\cdots,z_N)
\end{equation}
describing a probability density to find $p$ complex eigenvalues
around each of the points $z_1,\cdots,z_p$ while positions of the
remaining levels stay unobserved. Quite often, one is also
interested in the thermodynamic limit of the correlation function
\begin{eqnarray}
    \label{magnify}
    \rho_p(z_1,\cdots,z_p) = \lim_{N \rightarrow \infty}
    \frac{1}{\delta_N^{2p}}\, R_p \left(
    \frac{z_1}{\delta_N},\cdots,\frac{z_p}{\delta_N}; N
    \right)
\end{eqnarray}
that magnifies spectrum resolution on the appropriate energy scale
$\delta_N$ while letting the matrix size $N$ tend to infinity.

The multi-fold integral in (\ref{p-point}) can explicitly be
evaluated by adopting the Gaudin-Mehta\cite{M-1991} method of
orthogonal polynomials originally introduced in the context of
Hermitean random matrix theory. It is a straightforward exercise
to demonstrate that $R_p(z_1,\cdots,z_p;N)$ admits the determinant
representation
\begin{eqnarray}
    \label{Rp-det}
    R_p(z_1,\cdots,z_p;N) = \det \left[ K_N (z_k,\bar{z}_{\ell})
    \right]_{k,\ell=1,\cdots,p}
\end{eqnarray}
involving the scalar kernel
\begin{eqnarray}
    \label{KN-def}
    K_N(z,z^\prime) = w(z,\bar{z}) \, w(z^\prime, \bar{z}^\prime)
    \sum_{\ell=0}^{N-1} P_\ell(z) P_\ell(z^\prime) \nonumber
\end{eqnarray}
expressed in terms of polynomials $P_\ell(z)$ orthonormal in the
complex plane $z=x+iy$
\begin{eqnarray}
    \int d^2 Z \, w^2(z,\bar{z}) \, P_k(z) P_{\ell}(\bar{z}) = \delta_{k \,\ell}
\end{eqnarray}
with respect to the measure $w^2(z,\bar{z}) \, d^2 Z$.

For instance, it follows from (\ref{Rp-det}) that the density of
states and the two-point correlation function equal
\begin{eqnarray}
    \label{r1}
    R_1(z;N) = K_N(z,\bar{z})
\end{eqnarray}
and
\begin{equation}
    \label{r2}
    R_2(z_1,z_2;N) = K_N(z_1,\bar{z}_1) K_N(z_2,\bar{z}_2)
    - \left| K_N(z_1,\bar{z}_2) \right|^2,
\end{equation}
respectively.

For the Gaussian measure, the orthonormal polynomials $P_\ell(z)$
are just monomials
\begin{eqnarray}
    \label{OPs}
    P_\ell(z) = \frac{z^\ell}{\sqrt{\pi \Gamma(\ell+1)}}
\end{eqnarray}
leading to the scalar kernel
\begin{eqnarray}
    \label{KN}
    K_N(z,z^\prime) = \frac{1}{\pi} \,
    e^{-z\bar{z}/2}\, e^{-z^\prime\bar{z}^{\,\prime}/2}
    \sum_{\ell=0}^{N-1} \,
    \frac{(zz^\prime)^\ell}{\Gamma(\ell+1)}.
\end{eqnarray}

\textbf{\emph{Density of states.}}---Put into the integral form,
the kernel (\ref{KN}) yields the finite--$N$ density of states
\begin{eqnarray}
    \label{N-dos}
    R_1(z;N) = \frac{e^{-z\bar{z}}}{\pi \Gamma(N)}
    \, \int_0^\infty d\lambda \, e^{-\lambda}
    (\lambda+ z\bar{z})^{N-1}.
\end{eqnarray}
In terms of the upper incomplete gamma function
\begin{eqnarray}
    \label{Incomplete}
    \Gamma(a,x) = \int_x^\infty dt \, t^{a-1} e^{-t}
\end{eqnarray}
the level density equivalently reads
\begin{eqnarray}
    \label{N-dos-equiv}
    R_1(z;N) = \frac{\Gamma(N, z{\bar z})}{\pi \Gamma(N)}.
\end{eqnarray}
A careful analysis of the integral (\ref{N-dos}) shows that, in
the large--$N$ limit, $N$ complex eigenvalues are (almost)
uniformly distributed within a circle of the radius $\sqrt{N}$
centered at $z=0$,
\begin{eqnarray}
    \label{Gin-Circle}
    R_1(z;N \gg 1) \simeq \pi^{-1} \theta(\sqrt{N} - |z|),
\end{eqnarray}
$\theta(x)$ being a Heaviside step function.

In the vicinity $z_c=(\sqrt{N} + u)\, e^{i\varphi}$ of the edge
$|z|=\sqrt{N}$ of the two-dimensional eigenvalue support described
by (\ref{Gin-Circle}), the density of states sharply crosses over
from $R_1(z;N \gg 1)=\pi^{-1}$ at $|z| < \sqrt{N}$ to $R_1(z;N \gg
1)=0$ at $|z| > \sqrt{N}$. The crossover is described by the local
density of eigenvalues $R_1^{\rm{(tails)}}(u)=R_1(z_c;N)$ which,
in the large--$N$ limit, turns out to be independent of the matrix
size $N$,
\begin{eqnarray}
    \label{edge-crossover}
    R_1^{\rm{(tails)}}(u) = \frac{1}{\pi\,(2\pi)^{1/2}}
    \, \int_{2u}^{\infty} dt\, e^{-t^2/2}.
\end{eqnarray}
Expressed in terms of the complementary error function
\begin{eqnarray}
    \label{error-fun}
    {\rm erfc}(x) = \frac{2}{\sqrt{\pi}} \, \int_x^\infty dt
    \, e^{-t^2}
\end{eqnarray}
the local density of states reads
\begin{eqnarray}
    \label{edge-crossover-erfc}
    R_1^{\rm{(tails)}}(u) = \frac{1}{2\pi}\,
    {\rm erfc\,}\left(u\sqrt{2}\right).
\end{eqnarray}
At $|u| \gg 1$, the tail asymptotics can be read out of
(\ref{edge-crossover-erfc}) and are given by
\begin{eqnarray}
    \label{crossover-asymptot}
    R_1 ^{\rm{(tails)}}(u) \simeq \frac{1}{\pi} \,
    \left(
        \theta(-u) +  \frac{e^{-2u^2}}{2(2\pi)^{1/2} \,u}
    \right).
\end{eqnarray}

\textbf{\emph{Two-point correlation function.}}---At finite $N$,
this simplest fluctuation characteristics is given by (\ref{r2})
and (\ref{KN}). In the large--$N$ limit, when both $z_1$ and $z_2$
in (\ref{r2}) are situated inside the circle $|z| < \sqrt{N}$, the
kernel (\ref{KN}) reduces to simple exponentials so that the
two-point correlation function becomes $N$--independent,
\begin{eqnarray}
    \label{r2-bulk}
    R_2(z_1, z_2) = R_2(z_1, z_2; N \gg 1) = \frac{1}{\pi^2}\, \left(1 - e^{-|z_1-z_2|^2}\right).
\end{eqnarray}
Make notice that the two-point correlation function
$R_2(z_1,z_2;N)$ differs from the {\it density-density}
correlation function ${\hat R}(z_1,z_2;N)$ defined by
\begin{equation}
    \label{dos-dos}
    {\hat R}(z_1,z_2;N)= \left<
        {\rm tr}\, \delta^2(z_1 - {\cal H})\;
        {\rm tr}\, \delta^2(z_2 - {\cal H})
    \right>_{{\cal H} \in {\rm GinUE}_N}.
\end{equation}
The latter contains an additional $\delta$--function contribution
\cite{Remark-Delta}
\begin{eqnarray}
    \label{dos-dos-explicit}
    {\hat R}(z_1,z_2) = {\hat R}(z_1,z_2;N)= \frac{1}{\pi} \, \delta^2(z_1 - z_2) + R_2(z_1, z_2)
\end{eqnarray}
coming from the self-correlation of eigenlevels that chanced to
meet in the complex plane.

\section{Fermionic replica field theory}
\label{Sec.Fermions} In this section we are going to re-derive the
above nonperturbative results by mapping the Ginibre ensemble
(\ref{P-of-H}) of complex non-Hermitean random matrices onto a
zero-dimensional (0D) {\it fermionic} replica field theory. Since
the nonperturbative fluctuation formulae collected in Sec.
\ref{Sec.Mehta} are well known for almost four decades (we remind
that Ginibre's work \cite{G-1965} dates back to 1965), one may
wonder why we should bother ourselves with such a minor issue. The
answer prompted by the discussion in Sec. \ref{Sec.Intro} is
twofold. First, more than twenty years after their invention
\cite{W-1979,SW-1980}, replica field theories largely remain
unexplored territory from the viewpoint of their {\it
controllable} treatment away from a perturbative sector. Second,
learning intrinsic integrable structure of fermionic replica field
theories in the simplest 0D limit -- apart from encountering
indisputable mathematical beauty of exact theory -- creates a
basis for future work beyond the RMT: Given an intimate
connection\cite{K-2002} between integrability and the underlying
physical symmetries, one may hope that some crucial
characteristics of 0D replica partition functions are in fact not
so peculiar to these simple models but remain true in a more
general setting.

\subsection{Replica partition functions}
\label{SubSec.Partitions} \textbf{\emph{Density of states.}}---To
determine the average density of complex eigenvalues of the matrix
Hamiltonian ${\cal H} \in {\rm GinUE}_N$, we use a proper
modification\cite{NK-2002} of (\ref{r-limit}) and (\ref{replica}).
A new recipe \cite{NK-2002} is
\begin{eqnarray}
    \label{dos-def}
    R_1(z;N) = \lim_{n \rightarrow 0} \frac{1}{\pi n} \,
    \frac{\partial^2}{\partial z \partial \bar{z}}\,
    Z_n(z,\bar{z};N)
\end{eqnarray}
where the replica partition function $Z_n(z,\bar{z};N)$ is
determined by the matrix integral
\begin{equation}
    \label{Zn-before}
    Z_{n \in {\mathbb R}}(z, \bar{z};N) = \left<
    \det {}^n (z-{\cal H}) \det {}^n (\bar{z}-{\cal H}^\dagger)
    \right>_{{\cal H} \in {\rm GinUE}_N}.
\end{equation}
Representing each of the determinants in (\ref{Zn-before}) as a
field integral over an $N$--component fermionic field and
performing the averaging over ${\cal H} \in {\rm GinUE}_N$, one
maps $Z_n(z,\bar{z};N)$ onto a fermionic replica sigma model,
$Z_n(z,\bar{z};N) \mapsto {\tilde Z}_n(z,\bar{z};N)$, of the
form\cite{NK-2002}
\begin{eqnarray}
    \label{Zn-mapped}
    {\tilde Z}_{n \in {\mathbb Z}^+}(z, \bar{z};N) =
    \left<
    \det {}^N
    \left(
        \begin{array}{cc}
          z & -{\cal Q} \\
          {\cal Q}^\dagger & \bar{z} \\
        \end{array}
    \right)
    \right>_{{\cal Q} \in {\rm GinU}_n}.
\end{eqnarray}
Importantly, while $Z_n(z,\bar{z};N)$ in (\ref{Zn-before}) is
defined for arbitrary $n \in {\mathbb R}$, the replica parameter
$n$ in the representation (\ref{Zn-mapped}) for the mapped replica
partition function ${\tilde Z}_n(z,\bar{z};N)$ is restricted -- by
derivation -- to positive integers only, $n \in {\mathbb Z}^+$. To
emphasise this difference between two types of partition
functions, we will write either `tilded' ${\tilde Z}_n$ ($n \in
{\mathbb Z^+}$) or `untilded' $Z_n$ ($n \in {\mathbb R}$).

The matrix integral (\ref{Zn-mapped}) can be reduced to an
$n$--fold integral \cite{NK-2002} by making use of a singular
value decomposition of a complex matrix ${\cal Q} \in {\mathbb
C}^{n\times n}$. Expressing it as $Q=U\Lambda V$ where $U \in
\texttt{U}(n)/ \texttt{U}(1)^n$, $V \in  \texttt{U}(n)$ and
$\Lambda = {\rm diag}(\lambda_1^{1/2},\cdots,\lambda_n^{1/2})$
with $\lambda_\ell \ge 0$, and calculating a Jacobian of the
transformation ${\cal Q} \rightarrow (\Lambda,U,V)$, one derives
\cite{NK-2002}
\begin{equation}
    \label{Zn-mapped-eigen}
    {\tilde Z}_n(z,\bar{z};N) = \int_0^\infty
    \prod_{\ell=1}^n
    d\lambda_\ell \, e^{-\lambda_\ell}\, (\lambda_\ell +
    z\bar{z})^N \prod_{\ell_1 < \ell_2=1}^n |\lambda_{\ell_1} -
    \lambda_{\ell_2}|^2.
\end{equation}
We reiterate that this representation makes sense for $n \in
{\mathbb Z}^+$ only. This is precisely the reason why the replica
limit (\ref{dos-def}) with $Z_n$ replaced by ${\tilde Z}_n$ cannot
be implemented directly.

\textbf{\emph{Density-density correlation function.}}---Similarly
to the level density (\ref{dos-def}), the density-density
correlation function (\ref{dos-dos}) can be retrieved from the
replica limit
\begin{equation}
    \label{dd-replica}
    {\hat R}(z_1,z_2;N) = \lim_{n \rightarrow 0}\, \frac{1}{\pi^2 n^2}
    \, \frac{\partial^2}{\partial z_1 \partial \bar{z}_1}
    \frac{\partial^2}{\partial z_2 \partial \bar{z}_2}
    \, Z_n(z_1,{\bar z}_1; z_2,{\bar z}_2;N)
\end{equation}
where, for arbitrary $n \in {\mathbb R}$, the replica partition
function $Z_n(z_1,{\bar z}_1; z_2,{\bar z}_2;N)$ is defined by
\begin{eqnarray}
    \label{zn4}
    Z_{n \in {\mathbb R}}(z_1,{\bar z}_1; z_2,{\bar z}_2;N) =
    \left<
    \det {}^n (z_1-{\cal H}) \det {}^n (\bar{z}_1-{\cal H}^\dagger)
    \det {}^n (z_2-{\cal H}) \det {}^n (\bar{z}_2-{\cal
    H}^\dagger)
    \right>_{{\cal H} \in {\rm GinUE}_N}.
\end{eqnarray}
Upon a fermionic mapping \cite{NK-2002} which restricts the
replica parameter to $n \in {\mathbb Z}^+$, this generating
function can be rewritten as
\begin{equation}
    \label{Zndd-mapped}
    {\tilde Z}_{n \in {\mathbb Z}^+}(z_1, \bar{z}_1;z_2, \bar{z}_2;N) =
    \left< \det {}^N
    \left(
        \begin{array}{cc}
          {\cal Z} & -{\cal Q} \\
          {\cal Q}^\dagger & \bar{{\cal Z}} \\
        \end{array}
    \right)
    \right>_{{\cal Q} \in {\rm GinUE}_{2n}}
\end{equation}
with the diagonal matrix ${\cal Z}= {\rm diag}(z_1 \, \openone_n,
\; z_2 \, \openone_n)$. Note a tilde in (\ref{Zndd-mapped}).

For $z_1,z_2$ finite and of order unity, ${\tilde Z}_{n \in
{\mathbb Z}^+}$ can be reduced\cite{NK-2002} to a matrix integral
over $U\in \texttt{U}(2n)$ which eventually boils down to the
$n$--fold integral\cite{Remark-Zn}
\begin{eqnarray}
    \label{dd-nfold}
    {\tilde Z}_{n}(z_1, \bar{z}_1;z_2, \bar{z}_2;N) =
    e^{-2n(N-z\bar{z})}
    \int_{-1}^{+1}
    \prod_{\ell=1}^n
    d\lambda_\ell \, e^{(\omega\bar{\omega}/2)\lambda_\ell}
    \, \prod_{\ell_1 < \ell_2=1}^n |\lambda_{\ell_1} -
    \lambda_{\ell_2}|^2.
\end{eqnarray}
Here,
\begin{eqnarray}
    \label{wz-not}
    z = \frac{z_1+z_2}{2}, \;\; \omega = z_1 - z_2.
\end{eqnarray}
As is the case (\ref{Zn-mapped-eigen}), this representation makes
sense for $n \in {\mathbb Z}^+$ so that the replica limit
(\ref{dd-replica}) with $Z_n$ replaced by ${\tilde Z}_n$ cannot be
implemented directly.

\subsection{Replica partition functions as members of a positive Toda Lattice Hierarchy}
\label{Sub.Sec.Toda} By derivation, the $n$--fold integral
representations (\ref{Zn-mapped-eigen}) and (\ref{dd-nfold}) of
the replica partition functions ${\tilde Z}_n(z,\bar{z};N)$ and
${\tilde Z}_n(z_1,\bar{z}_1;z_2,\bar{z}_2;N)$ stay valid for $n
\in {\mathbb Z}^+$ only. Therefore, any attempt to retrieve
spectral fluctuation properties of the matrix Hamiltonian ${\cal
H}$ out of the replica limits (\ref{dos-def}) and
(\ref{dd-replica}) with $Z_n$ replaced by ${\tilde Z}_n$ will
inevitably face the problem of analytic
continuation\cite{Remark-Parisi} of ${\tilde Z}_n$'s away from $n$
positive integers. For this procedure to be controlled, an {\it
exact} result for ${\tilde Z}_n$ is desired to start with.

For approximate treatment\cite{KM-1999a,Z-1999} of the above
replica partition functions the reader is referred to Ref.
\cite{NK-2002}. In this subsection we wish to explore another
route which rests on {\it exact} and, therefore, truly
nonperturbative evaluation of replica partition functions. A
connection between nonlinear replica $\sigma$ models and the
theory of integrable hierarchies is at the heart of our formalism
\cite{K-2002}. A proof that the nonperturbative fermionic replica
partition functions form a positive, semi-infinite Toda Lattice
Hierarchy is the first important outcome of our approach.

\textbf{\emph{Bulk density of states.}}---To show how the Toda
Lattice Hierarchy emerges in the context of
(\ref{Zn-mapped-eigen}), we represent the Vandermonde determinant
there as
\begin{eqnarray}
    \label{Vandermonde}
    \prod_{\ell_1 < \ell_2=1}^n (\lambda_{\ell_1} - \lambda_{\ell_2})=
    \det(\lambda_k^{\ell-1})_{k,\ell =1,\cdots,n},
\end{eqnarray}
simultaneously shift all $\lambda_\ell$'s  therein by $z\bar{z}$,
and perform the $n$--fold integral (\ref{Zn-mapped-eigen}) by
means of the Andr\'eief--de Bruijn integration formula
\cite{A-1883,dB-1955}
\begin{eqnarray}
    \label{dB-formula}
    \int \prod_{\ell=1}^n d\mu(\lambda_\ell) \;
    \det [A_k(\lambda_\ell)]_{k,\ell=1,\cdots,n}
    \det [B_k(\lambda_\ell)]_{k,\ell=1,\cdots,n}
    = n! \; \det \left(
        \int d\mu(\lambda) A_k(\lambda) \, B_\ell (\lambda)
    \right)_{k,\ell=1,\cdots,n}\;
\end{eqnarray}
which holds for any benign integration measure $d\mu(\lambda)$
given convergence of the integrals involved. Up to (irrelevant for
our purposes) factorial prefactor, the interim result is
\begin{equation}
    \label{Zn-dos-1}
    {\tilde Z}_n(z,\bar{z};N) =
    \det \left(
        \int_0^{\infty} d\lambda \, e^{-\lambda}
        (\lambda+ z\bar{z})^{N+k+\ell}
    \right)_{k,\ell=0,\cdots,n-1}.
\end{equation}
While exhibiting some beauty (in particular, $e^{-z\bar{z}}{\tilde
Z}_1(z,\bar{z};N)$ coincides, up to a normalisation prefactor,
with the density of states $R_1(z;N+1)$ in ${\rm GinUE}_{N+1}$,
see (\ref{N-dos})), the representation (\ref{Zn-dos-1}) is not
very informative or helpful. What is more helpful is another
though totally equivalent form of (\ref{Zn-dos-1}),
\begin{eqnarray}
    \label{Zn-dos-2}
    {\tilde Z}_n(z,\bar{z};N) = e^{nz\bar{z}} (z\bar{z})^{n(n+N)}
    \, {\tilde \tau}_n(z\bar{z};N),
\end{eqnarray}
which involves the {\it Hankel determinant}
\begin{eqnarray}
    \label{dos-tau}
    {\tilde \tau}_n(z\bar{z};N) = \det \left[
        \partial_{(z\bar{z})}^{k+\ell} {\tilde \tau}_1(z\bar{z}; N)
    \right]_{k,\ell=0,\cdots,n-1}
\end{eqnarray}
with
\begin{eqnarray}
    \label{dos-tau-1}
    {\tilde \tau}_0 (z\bar{z}; N) = 1, \;\;\;\;
    {\tilde \tau}_1(z\bar{z}; N) = \int_1^{\infty} d\lambda \,
        \lambda^N e^{-z\bar{z} \lambda} =\frac{ \Gamma(N+1, z\bar{z})}
        {(z\bar{z})^{N+1}}.
\end{eqnarray}
Here, $\Gamma(a, x)$ is the upper incomplete gamma function
(\ref{Incomplete}). The initial condition for ${\tilde
\tau}_0(z\bar{z};N)$ reflects the fact that ${\tilde
Z}_0(z,\bar{z};N)=Z_0(z,\bar{z};N)=1$, see (\ref{Zn-before}).

The Hankel determinant (\ref{dos-tau}) is a remarkable object.
Whatever the function ${\tilde \tau}_1(z\bar{z};N)$ is, by virtue
of the Darboux Theorem \cite{D-1972}, the entire sequence
$\{{\tilde \tau}_{k \in {\mathbb Z^+}}\}$ satisfies the equation
\begin{eqnarray}
    \label{toda-eq}
    {\tilde \tau}_n \, {\tilde \tau}_n^{\prime\prime}
    - ({\tilde \tau}_n^{\prime})^2 = {\tilde \tau}_{n-1}
    {\tilde \tau}_{n+1}, \;\;\;\; n \in {\mathbb Z}^+
\end{eqnarray}
where the prime ${}^\prime$ stands for $\partial_{(z\bar{z})}$.
Equations (\ref{Zn-dos-2}) and (\ref{toda-eq}) taken together with
the known initial conditions ${\tilde \tau}_0=1$ and ${\tilde
\tau}_1$ given by (\ref{dos-tau-1}) establish\cite{K-2002} a
hierarchy between nonperturbative fermionic replica partition
functions ${\tilde Z}_{n}$ with different $n \in {\mathbb Z^+}$.
The {\it exact} result (\ref{Zn-dos-2}), (\ref{dos-tau-1}), and
(\ref{toda-eq}) is an alternative to an {\it approximate}
positive-integer--$n$ treatment of the very same replica partition
function presented in Ref. \cite{NK-2002}.

{\it Equation (\ref{toda-eq}), known as positive, semi-infinite
Toda Lattice equation in the theory of integrable hierarchies
\cite{T-2000}, is the first indication of exact solvability hidden
in replica field theories}. Importantly, emergence of the Toda
Lattice Hierarchy is eventually due to the $\beta=2$ Dyson's
symmetry of the fermionic replica field theory encoded into the
squared Vandermonde determinant in (\ref{Zn-mapped-eigen}).

\textbf{\emph{Density of states at the edge.}}---The above
symmetry argument ensures that the replica partition function for
the edge density of states in the ${\rm GinUE}_N$ will obey the
same Toda Lattice equation albeit with different initial
conditions. Close to the edge $|z|=\sqrt{N}$ (see discussion next
to (\ref{Gin-Circle})), one is interested in the large--$N$
replica partition function ${\tilde Z}_n(z,\bar{z};N)$ taken at
$z=(\sqrt{N} + u)\, e^{i\varphi}$ in the regime $|u|\ll \sqrt{N}$.
The latter, denoted as ${\tilde Z}_n^{{\rm (tails)}}(u)$, is
independent of the matrix size $N$ and equals
\begin{equation}
    \label{Zn-dos-tails}
    {\tilde Z}_n^{{\rm (tails)}}(u) = \int_0^\infty
    \prod_{\ell=1}^n dt_\ell \, e^{-t_\ell^2/2 - 2 u t_\ell}
    \prod_{\ell_1 < \ell_2=1}^n |t_{\ell_1} -
    t_{\ell_2}|^2.
\end{equation}
Irrelevant numeric prefactors were omitted.

Much in line with previous calculations, the $n$--fold integral
(\ref{Zn-dos-tails}) can be transformed into the Hankel
determinant form
\begin{eqnarray}
    \label{Hankel-tails-1}
    {\tilde Z}_n^{{\rm (tails)}}(u) = \det \left[
        \partial_u^{k+\ell} {\tilde Z}_1^{{\rm (tails)}}(u)
    \right]_{k,\ell=0,\cdots,n-1}
\end{eqnarray}
with
\begin{equation}
    \label{Hankel-tails-2}
    {\tilde Z}_0^{{\rm (tails)}}(u)=1, \;\;\;\;\;
    {\tilde Z}_1^{{\rm (tails)}}(u) = e^{2u^2} \, \int_{2u}^{\infty} dt \, e^{-t^2/2}
    =\sqrt{\frac{\pi}{2}} \, e^{2u^2} {\rm erfc}\left(u\sqrt{2}\right).
\end{equation}
As soon as ${\tilde Z}_0^{{\rm (tails)}}(u)=1$ (normalisation),
the Darboux Theorem\cite{D-1972} can be applied to conclude that
the entire sequence $\{{\tilde Z}^{{\rm (tails)}}_{k\in {\mathbb
Z}^+}\}$ of replica partition functions at the edge of the
two-dimensional eigenvalue support belongs to the positive,
semi-infinite Toda Lattice Hierarchy
\begin{eqnarray}
    \label{toda-eq-tails}
    {\tilde Z}_n^{{\rm (tails)}} \, {\tilde Z}_n^{{{\rm (tails)}}\; \prime\prime}
    - ({\tilde Z}_n^{{{\rm (tails)}}\; \prime})^2 = {\tilde Z}_{n-1}^{{\rm (tails)}}
    {\tilde Z}_{n+1}^{{\rm (tails)}}, \;\;\;\; n \in {\mathbb Z}^+
\end{eqnarray}
the prime ${}^\prime$ stands for $\partial_u$. Also, similarly to
the observation made below (\ref{Zn-dos-1}), we notice that, up to
a prefactor, $e^{-2u^2}{\tilde Z}_1^{\rm (tails)}(u)$ coincides
with the edge density of states $R_1^{{\rm (tails)}}(u)$ as given
by (\ref{edge-crossover}). We will comment on this later on.

\textbf{\emph{Bulk eigenvalue correlations.}}---By the same token,
the replica partition function (\ref{dd-nfold}) designed to
calculate the density-density correlation function via the replica
limit (\ref{dd-replica}) belongs to a Toda Lattice Hierarchy, too.
Separating $z$ and $\omega$--dependent pieces in (\ref{dd-nfold}),
one derives
\begin{eqnarray}
    \label{Zn-cor-factor}
    {\tilde Z}_{n}(z_1, \bar{z}_1;z_2, \bar{z}_2;N) &=&
    e^{-2n(N-z\bar{z})} {\tilde Z}_n(\omega{\bar \omega})
\end{eqnarray}
where ${\tilde Z}_n(\omega{\bar \omega})$ determined by the
$n$--fold integral in (\ref{dd-nfold}) can be cast into the Hankel
determinant form \cite{K-2001,K-2002}
\begin{eqnarray}
    \label{dd-Hankel}
    {\tilde Z}_n(\omega{\bar \omega}) =
    \det \left(
        \partial_{(\omega\bar{\omega})}^{k+\ell} \;
        \frac{\sinh(\omega\bar{\omega}/2)}{\omega\bar{\omega}/2}
    \right)_{k,\ell=0,\cdots,n-1}.
\end{eqnarray}
The positive, semi-infinite Toda Lattice equation for $\{ {\tilde
Z}_{k \in {\mathbb Z}^+}\}$ readily follows by virtue of the
Darboux Theorem.

\subsection{Replica partition functions and Painlev\'e transcendents}
While important for revealing integrability of the field theory, the
Toda Lattice equation for {\it fermionic} replica partition
functions ${\tilde Z}_{n \in {\mathbb Z}^+}$ is not much helpful in
performing the replica limit, if taken {\it alone}. Indeed, a
positive Toda Lattice equation gives no close expression for
${\tilde Z}_{n}$ as a function of $n \in {\mathbb Z^+}$ that would
facilitate an analytic continuation to $n \in {\mathbb R}$ in
general, and to the region $0 \le n \ll 1$ in particular. As was
recently shown by Splittorff and Verbaarschot
\cite{SV-2003,SV-2003new}, this difficulty can be circumvented if
one succeeds in gaining a complementary hierarchical information
about {\it bosonic} replica partition function. (The latter
satisfies a negative\cite{K-2003} Toda Lattice equation). Viable yet
surprisingly efficient\cite{SV-2003new} for random matrix models
describing interactionless stochastic systems, this route is
certainly unavailable for exact replica description of physical
systems in presence of interaction \cite{F-1983} which requires to
use {\it either} bosonic {\it or} fermionic field integrals in order
to properly accommodate quantum statistics of interacting species.

Considering exact replica treatment of disordered interacting
systems as a legitimate goal, it would be conceptually important
to not rely on such a complementary information. Fortunately, for
0D interactionless systems at hand, it is indeed possible.
Miraculously, the same Toda lattice equation governs the behaviour
of so-called $\tau$-functions arising in the Hamiltonian
formulation \cite{O-1,O-2,O-3,O-4} of the six Painlev\'e
transcendents (PI -- PVI), which are yet another fundamental
object in the theory of nonlinear integrable systems. Luckily, the
Painlev\'e equations being second order nonlinear differential
equations contain the hierarchy (or replica) index $n$ as a {\it
parameter}. As will be demonstrated below, this feature of
Painlev\'e equations makes them serve as a proper starting point
for constructing a consistent analytic continuation of
nonperturbative replica partition functions away from $n$
integers. {\it This Painlev\'e reduction further confirms exact
solvability of replica $\sigma$ models and assists \cite{K-2002}
performing the replica limit}.

\textbf{\emph{Bulk density of states.}}---Certainly being an
option, the aforementioned Toda to Painlev\'e
reduction\cite{FW-2002} is not the only way to arrive at the
sought Painlev\'e representation of the replica partition function
${\tilde Z}_n(z,\bar{z};N)$. An alternative approach would rest on
the observation that the $n$--fold integral
(\ref{Zn-mapped-eigen}) is essentially a {\it Fredholm
determinant}\cite{TW-1994a} associated with a gap formation
probability
\begin{equation}
    \label{gap-Laguerre}
    {\mathbb E}_n^{(0,z\bar{z})}(0;a) = \int_{z\bar{z}}^\infty
    \prod_{\ell=1}^n
    d\lambda_\ell \, e^{-\lambda_\ell}\, \lambda_\ell^a
    \prod_{\ell_1 < \ell_2=1}^n |\lambda_{\ell_1} -
    \lambda_{\ell_2}|^2
\end{equation}
within the interval $(0,z\bar{z})$ in the spectrum of an auxiliary
$n\times n$ Laguerre unitary ensemble. Celebrated
result\cite{TW-1994b} due to Tracy and Widom states that
\begin{eqnarray}
    \label{gap-PV}
    {\mathbb E}_n^{(0,z\bar{z})}(0;a) = \exp \left(
    \int_0^{z\bar{z}} dt \,
    \frac{\sigma_{\rm V}(t)}{t}
    \right)
\end{eqnarray}
where $\sigma_{\rm V}(t) = \sigma_n(t;a)$ is the fifth Painlev\'e
transcendent satisfying the Jimbo-Miwa-Okamoto form of the
Painlev\'e V equation \cite{O-2,JMMS-1980}
\begin{eqnarray}
    \label{PV-eq}
    (t \sigma^{\prime\prime}_{\rm V})^2 -  (a \sigma^\prime_{\rm V})^2
    - (\sigma_{\rm V} - t
    \sigma^\prime_{\rm V})
    \left[\sigma_{\rm V} - t \sigma^\prime_{\rm V} + 4 \sigma^\prime_{\rm V}
    \left(\sigma^\prime_{\rm V}+ n + \frac{a}{2}\right)\right] = 0
\end{eqnarray}
supplemented by the boundary condition \cite{F-book,Remark-BCs}
\begin{equation}
    \label{PV-bc}
    \left. \sigma_{\rm V}(t)\right|_{t\rightarrow +\infty} \sim
    - nt + an - \frac{an^2}{t} + {\cal O}(t^{-2}).
\end{equation}
We, thus, derive an exact Painlev\'e V representation of the
replica partition function ${\tilde Z}_n(z,{\bar z};N)$ in the
form
\begin{eqnarray}
    \label{Zn-gapped}
    {\tilde Z}_n(z,\bar{z};N) = e^{nz\bar{z}} \,
    \exp \left(
    \int_0^{z\bar{z}} dt \,
    \frac{\sigma_{\rm V}(t)}{t}
    \right)
\end{eqnarray}
where $\sigma_{\rm V}(t)=\sigma_n(t;a=N)$.

Note that (\ref{PV-eq}) -- (\ref{Zn-gapped}) contain the replica
index $n \in {\mathbb Z}^+$ as a {\it parameter}. Taken together
with the fact that the above Painlev\'e representation encodes
{\it all} hierarchical inter-relations between the replica
partition functions with various replica indices, it is very
tempting to conjecture that (\ref{Zn-gapped}), as it stands, holds
beyond $n \in {\mathbb Z}^+$ as well. Indeed, in Appendices A and
B, it is proven that (\ref{PV-eq}) -- (\ref{Zn-gapped}) stay valid
for generic real valued $n > -1$ so that the replica limit
(\ref{dos-def}) with $Z_n$ substituted by ${\tilde Z}_n$ can
safely be implemented.

To proceed, we expand a solution to (\ref{PV-eq}) around $n=0$.
Owing to the normalisation ${\tilde Z}_{0}=1$, the expansion
starts with a term linear in $n$,
\begin{eqnarray}
    \label{PV-expansion}
    \sigma_{\rm V}(t) = \sigma_n(t;a) = \sum_{p=1}^\infty \,
    n^p  f_p(t;a).
\end{eqnarray}
Only the first term of the above series is of our interest since
the replica limit relates the bulk density of states $R_1(z;N)$ to
the function $f_1(t;N)$ as
\begin{eqnarray}
    \label{R1-f1}
    R_1(z;N)=\pi^{-1}[1 + f_1^\prime(z\bar{z};N)].
\end{eqnarray}
Here $f_1(t;N)$ satisfies the differential equation
\begin{equation}
    \label{f1-eq-1}
    (tf_1^{\prime\prime})^2 -(f_1-tf_1^\prime)^2 -
    2N f_1^\prime (f_1-tf_1^\prime) - (N f_1^\prime)^2 = 0
\end{equation}
subject to the conservation constraint\cite{Remark-BCs}
\begin{eqnarray}
    \label{f1-conservation}
    \int_{\mathbb C} d^{\,2}z \, R_1(z;N) = \int_0^\infty dt \, [1 + f_1^\prime(t;N)] = N.
\end{eqnarray}
Identifying a complete square in (\ref{f1-eq-1}), we reduce the
latter to the Kummer differential equation
\begin{eqnarray}
    \label{f1-eq-2}
    f_1+ (N-t)f_1^\prime \pm tf_1^{\prime\prime}=0.
\end{eqnarray}
The constraint (\ref{f1-conservation}) makes us look for those
solutions $f_1(t;N)$ whose first derivative $f_1^\prime$ is
bounded at $t=+\infty$ and possibly has an integrable singularity
at $t=+0$. This class of functions sought welcomes the sign $(-)$
in (\ref{f1-eq-2}) leading to a general solution
\begin{equation}
    \label{f1-gen-sol}
    f_1(t;N) = C_1\, (N-t) + C_2\, t^{N+1} \; {}_1 F_1 (N, N+2;
    -t).
\end{equation}
The constraint ({\ref{f1-conservation}) uniquely fixes unknown
constants $C_1$ and $C_2$ be $C_1=0$ and $C_2= - 1/\Gamma(N+2)$.
This yields \cite{Remark-BCs-check}
\begin{eqnarray}
    \label{f1-eq-fin}
    f_1(t;N) = - \frac{t^{N+1}}{\Gamma(N+2)} \; {}_1
    F_1(N;N+2;-t)
\end{eqnarray}
where ${}_1 F_1(a;b;t)$ is the confluent hypergeometric function
of Kummer. Consequently, the replica limit (\ref{dos-def}) of the
fermionic partition function (\ref{Zn-gapped}) results in the
density of states (\ref{R1-f1}) of the form
\begin{eqnarray}
    \label{N-dos-replicas}
    R_1(z;N) = \frac{\Gamma(N,z\bar{z})}{\pi \Gamma(N)}
\end{eqnarray}
which is equivalent to (\ref{N-dos}) and (\ref{N-dos-equiv}). The
small--$n$ expansion of the fermionic replica partition function
\begin{widetext}
\begin{eqnarray}
    \label{Zn-dos-expansion}
    \ln \,{\tilde Z}_n(z,\bar{z}; N) = n(z\bar{z}) \left[
        1 + \frac{(z\bar{z})^{N}}{(N+1)\Gamma(N+2)}
        \; {}_2 F_2 (N,N+1; N+2,N+2; -z\bar{z})
    \right] + {\cal O}(n^2)
\end{eqnarray}
\end{widetext}
is behind the result (\ref{N-dos-replicas}). It should be stressed
that the finite--$N$ result (\ref{N-dos-replicas}) cannot be
produced by approximate treatment \cite{NK-2002} of replicas which
heavily relies on availability of a large parameter ($N \gg 1$) in
the integral representation (\ref{Zn-mapped-eigen}).

\textbf{\emph{Density of states at the edge.}}---With the replica
partition function ${\tilde Z}_n^{\rm (tails)}$ given by
(\ref{Zn-dos-tails}), the density of states at the edge $|z| =
\sqrt{N}$ of the two-dimensional eigenvalue support is determined
by the replica limit
\begin{eqnarray}
    \label{replica-tail}
    R_1^{\rm (tails)}(u) = \lim_{n \rightarrow 0} \frac{1}{\pi n}
    \, \frac{\partial^2}{\partial u^2} \, {\tilde Z}_n^{\rm (tails)}(u).
\end{eqnarray}
The partition function ${\tilde Z}_n^{\rm (tails)}$ can again be
viewed as a Fredholm determinant associated with a gap formation
probability
\begin{equation}
    \label{gap-Hermite}
    {\mathbb E}_n^{(u,\infty)}(0) = \int_{-\infty}^u
    \prod_{\ell=1}^n
    d\lambda_\ell \, e^{-\lambda_\ell^2}\,
    \prod_{\ell_1 < \ell_2=1}^n |\lambda_{\ell_1} -
    \lambda_{\ell_2}|^2
\end{equation}
within the interval $(u,\infty)$ in the spectrum of an auxiliary
$n\times n$ Gaussian Unitary Ensemble. In terms of the fourth
Painlev\'e transcendent\cite{O-3} $\sigma_{\rm IV}$, it
reads\cite{TW-1994a,FW-2001}
\begin{eqnarray}
    \label{gap-PIV}
    {\mathbb E}_n^{(u,\infty)}(0) = {\mathbb E}_n^{(0,\infty)}(0) \exp
    \left(
         \int_0^u dt \, \sigma_{\rm IV}(t)
    \right)
\end{eqnarray}
where $\sigma_{\rm IV}(t)= \sigma_n(t;a=0)$ satisfies the
Painlev\'e IV equation in the Jimbo-Miwa-Okamoto form
\begin{equation}
    \label{PIV-eq}
    (\sigma^{\prime\prime}_{\rm IV})^2 -  4(t \sigma^\prime_{\rm IV} -  \sigma_{\rm
    IV})^2
    + 4 \sigma^\prime_{\rm IV} (\sigma^\prime_{\rm IV}-2a)(\sigma^\prime_{\rm IV}+2n)=
    0
\end{equation}
subject to the boundary condition\cite{Remark-BCs}
\begin{eqnarray}
    \label{PIV-bc}
    \left. \sigma_{\rm IV}(t) \right|_{t \rightarrow -\infty}
    \sim - 2nt - \frac{n(a+n)}{t} +{\cal O}(t^{-3}).
\end{eqnarray}
In both (\ref{PIV-eq}) and (\ref{PIV-bc}) the parameter $a$ has to
be set to zero. The above equations result in the Painlev\'e IV
representation of the fermionic replica partition function
\begin{eqnarray}
    \label{Zn-tails-gapped}
    {\tilde Z}_n^{\rm (tails)} (u) = e^{2n u^2} \,
     \exp \left(
         \int_0^{-u \sqrt{2}} dt \, \sigma_{\rm IV}(t)
    \right)
\end{eqnarray}
which holds for $n \in {\mathbb Z^+}$.

To perform the replica limit (\ref{replica-tail}), we follow the
technology that led us to the small--$n$ expansion
(\ref{Zn-dos-expansion}). To this end we have to
assume\cite{Remark-Proof} that the Painlev\'e IV representation
(\ref{Zn-tails-gapped}) stays valid in a vicinity of $n=0$.
Writing down\cite{Remark-Expan}
\begin{eqnarray}
    \label{PIV-expansion}
    \sigma_{\rm IV}(t) = \sigma_n(t;a) = \sum_{p=1}^\infty \,
    n^p  g_p(t;a),
\end{eqnarray}
one derives from here, (\ref{replica-tail}) and
(\ref{Zn-tails-gapped}) that
\begin{eqnarray}
    \label{R1-g1}
    R_1^{\rm (tails)}(u) = \frac{4}{\pi} \left[
    1 + \frac{1}{2}\, g_1^\prime (-u\sqrt{2};0)
    \right].
\end{eqnarray}
Equation for $g_1=g_1(t;a=0)$ follows from (\ref{PIV-eq}) and
(\ref{PIV-expansion}),
\begin{eqnarray}
    \label{g1-eq}
    g_1^{\prime\prime} \pm 2(t g_1^\prime -g_1) = 0
\end{eqnarray}
while the boundary conditions are \cite{Remark-BCs-g1}
\begin{eqnarray}
    \label{g1-prime-prop}
    \begin{array}{rl}
      g_1^\prime(-\infty)+2=0 & \;\;\;{\rm (convergence)} \\
      g_1^\prime(+\infty)+3/2=0 & \;\;\; {\rm (conservation)} \\
    \end{array}
\end{eqnarray}
The sign $(-)$ in (\ref{g1-eq}) leads to a solution with unbounded
first derivative $g_1^\prime(t)$ at both infinities and is
therefore incompatible with (\ref{g1-prime-prop}). Equation
(\ref{g1-eq}) with the sign $(+)$ yields a general solution
\begin{eqnarray}
    \label{g1-sol-plus}
    g_1(t) = C_1 \, t + C_2 \left(
        t\, {\rm erf}\,t + \frac{1}{\sqrt{\pi}}  \, e^{-t^2}
    \right).
\end{eqnarray}
Boundary conditions (\ref{g1-prime-prop}) fix the constants
\cite{Remark-BCs-tails} be $C_1=-7/4$ and $C_2=1/4$. By virtue of
(\ref{R1-g1}) this results in the density of states
\begin{eqnarray}
    \label{dos-tails}
    R_1^{\rm (tails)}(u) = \frac{1}{2\pi} \, {\rm erfc}\left( u\sqrt{2}
    \right).
\end{eqnarray}
This is identically equivalent to (\ref{crossover-asymptot}). The
small--$n$ expansion of the fermionic replica partition function
\begin{widetext}
\begin{eqnarray}
    \label{Zn-tails-expansion}
    \ln \, {\tilde Z}_n^{\rm (tails)}(u) = \frac{n}{4}
    \left[\,
    u^2  -
    \frac{1}{\sqrt{2\pi}} \, u e^{-2u^2}
    -\,\left( u^2 + \frac{1}{4}\right) \, {\rm erf}\left( u \sqrt{2}\right)
    \right] + {\cal O}(n^2)
\end{eqnarray}
is behind the result (\ref{dos-tails}). Being exact, the formula
(\ref{dos-tails}) describes the tails of level density {\it both}
inside $(u<0)$ and outside $(u>0)$ of the circle \cite{G-1965}
$|z|=\sqrt{N}$. The approximate treatment \cite{NK-2002} of
replicas has failed to reproduce the density of states outside the
circle, $|z| > \sqrt{N}$.

\textbf{\emph{Density-density correlation function.}} To determine
this spectral characteristics, we put the replica limit
(\ref{dd-replica}) into the form
\begin{eqnarray}
    \label{dd-replica-zw}
    {\hat R}(z_1,z_2) = \lim_{n \rightarrow 0}\, \frac{1}{\pi^2 n^2}
    \,
        \left( \frac{1}{4}\, \partial_z^2
        -
        \partial_\omega^2
        \right)
        \left( \frac{1}{4}\, \partial_{\bar z}^2
        -
        \partial_{\bar \omega}^2
        \right)
    \, {\tilde \Upsilon}_n(z,{\bar z}; \omega,{\bar \omega})
\end{eqnarray}
\end{widetext}
involving the variables $z$ and $\omega$ as defined by
(\ref{wz-not}). The notation ${\tilde \Upsilon}_n(z,{\bar z};
\omega,{\bar \omega})$ stands for the replica partition
function\cite{Remark-Zn}
\begin{equation}
    \label{ups}
    {\tilde \Upsilon}_n(z,{\bar z}; \omega,{\bar \omega}) =
    {\tilde Z}_n\left(z+\frac{\omega}{2}, {\bar z}+ \frac{\bar \omega}{2}; z-\frac{\omega}{2},
    {\bar z} -  \frac{\bar \omega}{2} \right).
\end{equation}
At $z \pm \omega/2$ of order unity, the $n$--fold integral
representation (\ref{dd-nfold}) makes it possible to express
${\tilde \Upsilon}_n$ in terms of a gap formation probability
\begin{equation}
    \label{gap-Laguerre-sup}
    {\mathbb E}_n^{(\omega {\bar \omega},\infty)}(0;a)
    =
    \int_0^{\omega \bar{\omega}}
    \prod_{\ell=1}^n
    d\lambda_\ell \, e^{-\lambda_\ell}\, \lambda_\ell^a
    \prod_{\ell_1 < \ell_2=1}^n |\lambda_{\ell_1} -
    \lambda_{\ell_2}|^2
\end{equation}
within the interval $(\omega{\bar \omega},\infty)$ in the spectrum
of an auxiliary $n \times n$ Laguerre unitary ensemble (compare to
(\ref{gap-Laguerre})). The result due to Tracy and
Widom\cite{TW-1994b} states that
\begin{eqnarray}
    \label{En-fredholm}
    {\mathbb E}_n^{(\omega {\bar \omega}, \infty)}(0,0) =
    \exp\left(
        - \int_{\omega {\bar \omega}}^\infty dt \,
        \frac{\sigma_{\rm V}(t)}{t}
    \right)
\end{eqnarray}
where the fifth Painlev\'e transcendent $\sigma_{\rm V}(t) =
\sigma_n(t;a=0)$ satisfies the equation (\ref{PV-eq}) with $a=0$
and meets the boundary condition \cite{B-2002}
\begin{eqnarray}
    \label{PV-bc2}
    \left. \sigma_{\rm V}(t) \right|_{t\rightarrow \infty}
    \sim
    \frac{t^{2n-1}}{\Gamma^2(n)}\, e^{-t}
    \left( 1 + {\cal O}(t^{-1}) \right).
\end{eqnarray}
The above equations yield, for $n \in {\mathbb Z^+}$, the exact
representation
\begin{equation}
    \label{Yn-E}
    {\tilde \Upsilon}_n(z,{\bar z}; \omega,{\bar \omega}) =
    \frac{\exp \left[ n(2 z{\bar z} + \omega {\bar \omega}/2 )\right]}
    {(\omega {\bar \omega})^{n^2}} \,
    \exp\left(
        - \int_{\omega {\bar \omega}}^\infty dt \,
        \frac{\sigma_{\rm V}(t)}{t}
    \right).
\end{equation}

To implement the replica limit, we have to analytically continue
(\ref{Yn-E}) into a vicinity of $n=0$. Although at the moment we
do not have a proof that (\ref{Yn-E}) as it stands also holds for
$n$ away from positive integers,  armed with the previous
experience we are going to conjecture that this is indeed the case
so that
\begin{eqnarray}
    \label{dos-dos-aux-1}
    \hspace{-0.5cm}
    {\hat R}(z_1,z_2) &=& \frac{1}{2 \pi^2} + \frac{1}{\pi^2}
    \lim_{n \rightarrow 0}
    \frac{1}{n^2} \left( \partial_\omega \partial_{\bar \omega}
    \right)^2
    \, \frac{\exp(n \omega {\bar \omega}/2)}{(\omega {\bar \omega})^{n^2}}
    \,
    \exp\left(
        - \int_{\omega {\bar \omega}}^\infty dt \,
        \frac{\sigma_{\rm V}(t)}{t}
    \right).
\end{eqnarray}
Equation (\ref{dos-dos-aux-1}) suggests that the two functions,
$h_1(t)$ and $h_2(t)$, of the small--$n$ expansion
\begin{eqnarray}
    \label{kappa-n-expansion}
    \sigma_{\rm V}(t) = \sum_{p=1}^\infty n^p \, h_p(t)
\end{eqnarray}
contribute the density-density correlation function in the replica
limit. As we tend to avoid explicit reference to the boundary
conditions for Painlev\'e transcendents, the easiest way to
determine $h_1$ is to notice that, at $|z| \ll \sqrt{N}$, the bulk
density of states equals
\begin{equation}
    \label{dos-another-way}
    R_1(z;N) = \lim_{n\rightarrow 0} \frac{1}{\pi n}
    \, \frac{\partial^2}{\partial z \partial {\bar z}}
    \, {\tilde Z}_n(z,{\bar z};0,0;N).
\end{equation}
This is so because, at $n \rightarrow 0$, the partition functions
(\ref{Zn-before}) and (\ref{zn4}) taken at $z_1=z$ and $z_2=0$
become indistinguishable if considered as functions of the energy
variable $z$. Given (\ref{ups}) and (\ref{Yn-E}), one derives
\begin{equation}
    \label{Zn-fredholm}
    {\tilde Z}_n(z,{\bar z};0,0;N) = \frac{\exp(nz{\bar z})}{(z {\bar z})^{n^2}} \exp
    \left(
        -\int_{z {\bar z}}^\infty dt \, \frac{\sigma_{\rm V}(t)}{t}
    \right).
\end{equation}
Only linear in $n$ term of the expansion (\ref{kappa-n-expansion})
contributes the replica limit (\ref{dos-another-way}) yielding
\begin{eqnarray}
    \label{dos-h1}
    R_1(z) = \frac{1 +
        h_1^\prime (z {\bar z})}{\pi}.
\end{eqnarray}
According to the replica result (\ref{N-dos-replicas}), this must
be equal to $1/\pi$ whence we conclude that the function $h_1(t)$
is a constant. Further, the equation
\begin{eqnarray}
    \label{h1-eq}
    (t h_1^{\prime \prime})^2 - (h_1 - t h_1^{\prime})^2=0
\end{eqnarray}
following from (\ref{PV-eq}) and (\ref{kappa-n-expansion}) sets
\begin{eqnarray}
    \label{h1-0}
    h_1(t) = 0.
\end{eqnarray}
Therefore, the first nontrivial term in the expansion
(\ref{kappa-n-expansion}) actually starts with $n^2 h_2(t)$ where
$h_2(t)$ satisfies the equation
\begin{eqnarray}
    \label{h2-eq}
    t h_2^{\prime \prime} \pm  ( t h_2^{\prime} - h_2) =0.
\end{eqnarray}
The sign $(-)$ is the one that meets existence arguments applied
to (\ref{En-fredholm}). As a result, we come down to
\begin{eqnarray}
    h_2(t) = C_1 t + C_2 \, E_2(t)
\end{eqnarray}
where $E_2(t)$ is the exponential integral
\begin{eqnarray}
    \label{EE}
    E_n(z) = \int_1^\infty dt \, \frac{e^{-zt}}{t^n}, \;\;\; {\rm
    Re \,}z > 0.
\end{eqnarray}
Again, by existence arguments, $C_1$ must be set to zero to ensure
convergence of the integral in the exponent of
(\ref{En-fredholm}). To fix the constant $C_2$, we make use of the
observation that ${\tilde {\Upsilon}}_n$ has to be finite at
$\omega=0$. This brings $C_2=1$ so that \cite{Remark-BCs-R}
\begin{eqnarray}
    \label{h2-final}
    h_2(t) = E_2(t).
\end{eqnarray}
Collecting (\ref{Yn-E}), (\ref{kappa-n-expansion}), (\ref{h1-0})
and (\ref{h2-final}), we end up with the following nonperturbative
small--$n$ expansion\cite{K-2002} of the logarithm of the replica
partition function:
\begin{widetext}
\begin{eqnarray}
    \label{ln-Zn}
    \ln \, {\tilde {\Upsilon}}_n(z,{\bar z}; \omega, {\bar \omega})
    =
    n \left(
        2z{\bar z} + \frac{\omega {\bar \omega}}{2}
    \right)
    - n^2 \big[
        \ln \, (\omega {\bar \omega})
        + E_1 (\omega {\bar \omega})
        - E_2 (\omega {\bar \omega})
    \big] + {\cal O}(n^3).
\end{eqnarray}
\end{widetext}
The replica limit (\ref{dd-replica-zw}) applied to (\ref{ln-Zn})
culminates in the exact result for the density-density correlation
function
\begin{eqnarray}
    \label{R2-final}
    {\hat R}(z_1, z_2) = \frac{1}{\pi} \delta^2(z_1-z_2)
    + \frac{1}{\pi^2}
    \left(
        1 - e^{-|z_1-z_2|^2}
    \right).
\end{eqnarray}
Notice a presence of the $\delta$--functional contribution in
(\ref{R2-final}) describing the self-correlation of complex
eigenlevels. The latter is inaccessible by the approximate
treatment \cite{NK-2002} of replicas.

\subsection{Comment on a puzzle}
In the paper\cite{NK-2002}, Nishigaki and Kamenev have noticed
that there exists a `striking resemblance' between the finite--$N$
densities of states and the replica partition functions for
non-Hermitean random matrices belonging to all three universality
classes $\beta=1,2$ and $4$. In the context of the present study,
the observation of the authors of Ref. \cite{NK-2002} can be
translated into the identity (see two remarks below
(\ref{Zn-dos-1}) and (\ref{toda-eq-tails}), respectively)
\begin{eqnarray}
    \label{identity-0}
    R_1(z; N) \propto e^{-z{\bar z}} \, {\tilde Z}_1(z,{\bar z}; N-1)
\end{eqnarray}
that links the density of states $R_1(z; N)$ in ${\rm GinUE}_N$ to
the replica partition function ${\tilde Z}_1(z,{\bar z}; N-1)$ for
the same ensemble albeit of a smaller dimension ${\rm
GinUE}_{N-1}$.

The identity (\ref{identity-0}) is not a miracle and can well be
understood as a consequence of the relation (\ref{r1}). Indeed,
viewing the scalar kernel $K_N(z_1,z_2)$ in (\ref{r1}) as a matrix
integral\cite{Remark-ZJ,ZJ-1998}
\begin{eqnarray}
    \label{Kernel-M-Integral}
    K_{N}(z_1, z_2) \propto
    e^{- z_1 {\bar z_1}/2}
    e^{- z_2{\bar z_2}/2}   \left<
    \det\, (z_1 - {\cal H}) \, \det\, (z_2 - {\cal H}^\dagger)
    \right>_{{\cal H} \in {\rm GinUE}_{N-1}} \nonumber
\end{eqnarray}
one readily identifies
\begin{eqnarray}
    \label{Kernel-Z-1}
    K_N(z,{\bar z}) \propto e^{-z{\bar z}} \, {\tilde Z}_1(z,{\bar z};N-1)
\end{eqnarray}
whence (\ref{identity-0}) follows\cite{Remark-GUE}. Interestingly,
so explained identity (\ref{identity-0}) taken together with
(\ref{dos-def}) leads, in the context of fermionic replica field
theory, to a much less trivial statement
\begin{equation}
    \label{claim-1}
    \frac{\partial^2}{\partial z \partial {\bar z}}
    \, {\tilde Z}_n(z,{\bar z};N) = \pi n \, e^{-z{\bar z}}
    {\tilde Z}_1(z,\bar{z};N-1) + {\cal O}(n^2).
\end{equation}
Similar small--$n$ expansions should exist for two other
($\beta=1$ and $4$) universality classes in non-Hermitean RMT.

\section{Discussion}
In the present paper we have offered a detailed account of a
nonperturbative approach\cite{K-2002} to zero-dimensional
fermionic replica field theories which is based on exact
representation of replica partition functions in terms of
Painlev\'e transcendents. Focussing on Ginibre ensemble
\cite{G-1965} of complex non-Hermitean random matrices, we have
revealed an intrinsic integrability of associated replica field
theories. It materialises in two ways: First, at $n \in {\mathbb
Z}^+$, the replica partition functions were proven to belong to a
positive, semi-infinite Toda Lattice Hierarchy. Second, the very
same replica partition functions were shown to be expressible in
terms of solutions to Painlev\'e equations which (i) contain the
replica index as a single {\it parameter} and which (ii)
implicitly encode {\it all} hierarchical inter-relations between
the fermionic replica partition functions with various replica
indices.

The above two observations [(i) and (ii)] led us to conjecture
that Painlev\'e representations of fermionic replica partition
functions stay valid beyond $n \in {\mathbb Z}^+$ and, in
particular, in a vicinity of $n=0$. Indeed, for a particular case
of the replica partition function ${\tilde Z}_n(z,{\bar z};N)$
designed to determine the bulk density of complex eigenvalues,
this conjecture was rigorously proven by appealing to the Okamoto
$\tau$-function theory of the Painlev\'e V. (Similar, in spirit,
proof was previously given \cite{K-2002} in the context of the
one-point Green function in the finite--$N$ Gaussian Unitary
Ensemble where a Painlev\'e IV equation arises). Once justified,
it is no surprise that taking the replica limit of the
Painlev\'e-represented fermionic partition function ${\tilde
Z}_n(z,{\bar z};N)$ has culminated in reproducing exact
nonperturbative results for the bulk density of states. In other
cases, which include the tails of level density and the
density-density correlation function in the spectrum bulk,
although implemented without a formal justification, the replica
limit of replica partition functions expressed in terms of
Painlev\'e transcendents has also brought exact nonperturbative
results. This fact as well as other encouraging applications
\cite{K-2002} of the present method make us look further into the
rational reasons behind its success. Possibly, recent developments
\cite{CDZ-2003} in the field of extended Toda Hierarchy may give
us the right lead.

Finally, a remark is in order aimed to pinpoint the difference
between the approach \cite{K-2002} detailed in this paper and a
complementary approach recently elaborated in Refs.
\cite{SV-2003,SV-2003new}. Although both approaches exploit the very
same replica representations of quantum correlation functions as a
starting point, the two frameworks are conceptually different. The
present approach\cite{K-2002,K-2003} based on exact Painlev\'e
evaluation of {\it fermionic} replica partition functions followed
by their continuation into a vicinity of $n=0$ makes {\it no}
reference whatsoever to {\it bosonic} partition functions. On the
contrary, the approach\cite{SV-2003,SV-2003new} exploiting the
replica limit of the Toda Lattice equation for replica partition
functions rests explicitly, and unavoidably, on the observation that
suitably normalised fermionic and bosonic replica partition
functions are the members of a {\it single} Toda Lattice Hierarchy
albeit belonging to its positive (fermionic) and negative (bosonic)
branches, respectively. Implemented on the level of such an infinite
-- {\it supersymmetric in essence} -- Toda Lattice equation, the
replica limit reveals a remarkable factorisation of quantum
correlation functions for an interactionless matrix Hamiltonian into
a product of {\it both} fermionic and bosonic partition functions.
It is this factorisation which -- in order to be materialised on the
operational level -- explicitly infuses\cite{New-Remark,Jac-Comment}
a missing bosonic (or fermionic) information to what early appeared
to be a pure fermionic (or bosonic) formulation of the field theory.
While facilitating calculation of quantum correlation functions in
the interactionless case, this feature makes the approach
\cite{SV-2003,SV-2003new} be potentially inapplicable for an exact
nonperturbative replica treatment\cite{F-1983} of 0D Hamiltonians
with interactions\cite{KAA-2000,OBWH-2001,ABG-2002} whose presence
requires using of {\it either} bosonic {\it or} fermionic field
integrals in order to accommodate a proper quantum statistics of
interacting species.

\begin{acknowledgments}
I thank Alex Kamenev and Jac Verbaarschot for clarifying
correspondence and Joshua Feinberg, Shmuel Fishman and Ady Stern for
helpful discussions. This work was supported in part by the Albert
Einstein Minerva Centre for Theoretical Physics at the Weizmann
Institute of Science, and by the Israel Science Foundation founded
by the Israel Academy of Sciences and Humanities, through the grant
No 286/04.

\end{acknowledgments}

\vspace{0.5cm}
\appendix
\section{Replica partition function $Z_n(z,{\bar z})$ for the
density of states beyond $n \in {\mathbb Z^+}$}

To prove that the Painleve representation (\ref{Zn-gapped}),
(\ref{gap-PV}), (\ref{PV-eq}) and (\ref{PV-bc}) stays valid for
generic real-valued $n >-1$, we start with the definition of the
replica partition function (\ref{Zn-before}). Upon Schur
decomposition of the matrix ${\cal H} \in {\rm GinUE}_N$ specified
below (\ref{P-of-H}), the partition function (\ref{Zn-before}) can
be put into the form
\begin{equation}
    \label{Zn-eigen}
    Z_n(z, {\bar z};N) = \int
    \prod_{\ell=1}^N
    d^2 Z_\ell \, e^{-z_\ell {\bar z}_{\ell}}\, |z_\ell -
    z|^{2n} \prod_{\ell_1 < \ell_2=1}^N |z_{\ell_1} -
    z_{\ell_2}|^2.
\end{equation}
Having introduced a new set of integration variables $\xi_\ell =
z_\ell - z$, we use (\ref{Vandermonde}) and a proper extension of
the Andr\'eief--de Bruijn formula (\ref{dB-formula}) into the
complex plane to derive
\begin{equation}
    \label{Zn-det-1}
    Z_n(z,\bar{z};N) =
    \det \left(
    \int d^2 \xi \; \xi^{n+k} \,{\bar \xi}^{n+\ell}
    \, e^{-(\xi + z)({\bar \xi}+{\bar z})}
    \right)_{k,\ell=0,\cdots,N-1}.
\end{equation}
Decomposition $\xi = \rho e^{i\theta}$ helps us perform
integration in (\ref{Zn-det-1}) eventually resulting, for $n>-1$,
in
\begin{widetext}
\begin{eqnarray}
    \label{Zn-det-2}
    Z_n(z, {\bar z};N) =
    e^{-N z{\bar z}}
    \det \left(
        \int_0^\infty d\rho \, e^{-\rho^2}
        \rho^{2n + 1 + k+\ell}
        I_{k-\ell} (2 |z| \rho)
    \right)_{k,\ell=0,\cdots,N-1}.
\end{eqnarray}
\end{widetext}
Here, $I_k(z)$ is the modified Bessel function. No care was taken
of numeric prefactors turning to unity in the replica limit $n
\rightarrow 0$.

The $(k,\ell)$ matrix entry in (\ref{Zn-det-2}) can be recognised
as the confluent hypergeometric function of Kummer
\begin{equation}
    \label{Zn-det-entry}
    \frac{1}{2}
    \, |z|^{k-\ell} \, \frac{\Gamma(n+k+1)}{\Gamma(k-\ell+1)}
    \; {}_1  F_1 (n+k+1; k-\ell+1; |z|^2).
\end{equation}
Applying the transformation formula
\begin{eqnarray}
    \label{tf}
    {}_1  F_1 (a;b;z) = e^z \, {}_1  F_1(b-a;b;-z)
\end{eqnarray}
and performing trivial row and column operations under the sign of
determinant, one concludes that
\begin{eqnarray}
    \label{Zn-det-3}
    Z_n(z, {\bar z};N) =
    \det \left[ L_{n+\ell}^{(k-\ell)} (- |z|^2)
    \right]_{k,\ell=0,\cdots,N-1}
\end{eqnarray}
where
\begin{eqnarray}
    \label{Laguerre}
    L_\nu^{(\lambda)}(- t) = \frac{\Gamma(\nu+\lambda+1)}{\Gamma(\nu+1)\Gamma(\lambda+1)} \,
    {}_1  F_1 (-\nu;\lambda+1;-t)
\end{eqnarray}
is the generalised Laguerre function.

Importantly, the representation (\ref{Zn-det-3}) holds for generic
real-valued $n>-1$. Moreover, as is proven in the Appendix B, the
determinant (\ref{Zn-det-3}) is identically equivalent to another
one
\begin{eqnarray}
    \label{Zn-det-4}
    Z^{(\tau)}_n(z, {\bar z};N) =
    \det \left[ L_{n+\ell-k}^{(k)} (- |z|^2)
    \right]_{k,\ell=0,\cdots,N-1}
\end{eqnarray}
arising in the $\tau$-function theory\cite{O-2,FW-2002} of the
fifth Painlev\'e transcendent so that
\begin{eqnarray}
    \label{Z-equiv}
    Z_n(z,{\bar z};N) = Z_n^{(\tau)}(z,{\bar z};N).
\end{eqnarray}

In the notation of Forrester and Witte \cite{FW-2002} (see their
Sec. 3.5) the latter can be expressed as
\begin{eqnarray}
    \label{Zn-det-5}
    Z_n^{(\tau)}(z, {\bar z};N) =  e^{-Nz{\bar z}} \, (z\bar{z})^{-N^2/2} \,
    {\bar \tau}[N](z{\bar z})
\end{eqnarray}
where ${\bar \tau}[N](t)$ is the Hankel determinant
\begin{equation}
    \label{Zn-det-6}
    {\bar \tau}[N](t) = \det
    \left[
    \delta_t^{k+\ell}
    \left( \, e^t \, {}_1 F_1 (-n; 1; -t)
    \right)
    \right]_{k,\ell=0,\cdots,N-1}
\end{equation}
built on the operator $\delta_t = t d/dt$.

In accordance with the Okamoto $\tau$-function theory
\cite{O-2,FW-2002} of the Painlev\'e V, the ${\bar \tau}[N](t)$
(which also depends on $n$) can be related to the fifth Painlev\'e
transcendent as \cite{FW-2002}
\begin{eqnarray}
    \label{sigma-V-tau}
    {\tilde \sigma}_{\rm V}(t) = t\frac{d}{dt} \,
    \left[  \,
        \ln \left(
            \frac{e^{-(n+N)t}}{t^{N^2/2}}
            \, {\bar \tau}[N](t)
        \right)
    \right].
\end{eqnarray}
Here, ${\tilde \sigma}_{\rm V}(t)=\sigma_{n}(t; N)$ obeys
\cite{FW-2002} the Painlev\'e V equation (\ref{PV-eq}) with $a = N$
subject\cite{F-book} to the boundary condition (\ref{PV-bc}). This
fact taken together with the equations (\ref{Zn-det-5}) and
(\ref{sigma-V-tau}) proves that the representation (\ref{PV-eq}) --
(\ref{Zn-gapped}), as it stands, stays valid for real valued $n
>-1$ in general and, in a vicinity of $n=0$, in particular.
This justifies taking the replica limit in the way described
between (\ref{PV-expansion}) and (\ref{N-dos-replicas}).

\section{A proof of equivalence of the two determinants (\ref{Zn-det-3}) and (\ref{Zn-det-4})}
Throughout the Appendix, $t$ denotes $t=z{\bar z}$.

(i) First, we use the identity
\begin{eqnarray}
    \label{id-1}
    \left(\frac{\partial}{\partial w}\right)^m \, L_\nu^{(\lambda)}
    (w) = (-1)^m \, L_{\nu-m}^{(\lambda+m)} (w)
\end{eqnarray}
to reduce (\ref{Zn-det-3}) to
\begin{eqnarray}
    \label{Zn-det-3a}
    Z_n(z, {\bar z};N) =
    \det \left[
    \left(\frac{\partial}{\partial t}\right)^k
    L_{n+\ell+k}^{(-\ell)} (- t)
    \right]_{k,\ell=0,\cdots,N-1}.
\end{eqnarray}
(ii) Second, we use the identity
\begin{eqnarray}
    \label{id-2}
    \left(\frac{\partial}{\partial w}\right)^m \,
    \left[
        e^{-w} w^{\lambda}\, L_\nu^{(\lambda)}
        (w)
    \right]
    =  \frac{\Gamma(\nu+m+1)}{\Gamma(m+1)} \,
    e^{-w} w^{\lambda-m}
    \, L_{\nu+m}^{(\lambda-m)}
    (w)
\end{eqnarray}
to bring (\ref{Zn-det-3a}) to
\begin{eqnarray}
    \label{Zn-det-3b}
    Z_n(z, {\bar z};N) =
    \det \left[
    \left(\frac{\partial}{\partial t}\right)^k
    \left( t^\ell L_{n}^{(\ell)} (- t) \right)
    \right]_{k,\ell=0,\cdots,N-1}.
\end{eqnarray}
(iii) Third, we use the identity
\begin{eqnarray}
    \label{id-3}
    L_\nu^{(-m)}
        (w) =  \frac{w^m}{(-\nu)_m} \;
        L_{\nu-m}^{(m)}
    (w)
\end{eqnarray}
to rewrite (\ref{Zn-det-3b}) as
\begin{eqnarray}
    \label{Zn-det-3c}
    Z_n(z, {\bar z};N) =
    \det \left[
    \left(\frac{\partial}{\partial t}\right)^k
    L_{n+\ell}^{(-\ell)} (- t)
    \right]_{k,\ell=0,\cdots,N-1}
\end{eqnarray}
or, equivalently, as
\begin{eqnarray}
    \label{Zn-det-3d}
    Z_n(z, {\bar z};N) =
    \det \left[
    L_{n+\ell-k}^{(k-\ell)} (- t)
    \right]_{k,\ell=0,\cdots,N-1}.
\end{eqnarray}
The transformations (i) to (iii) were aimed at expicit elimination
of deep cancellations occurring in the determinant
(\ref{Zn-det-3}) as can be recognised by comparing the entries of
matrices under the sign of determinant in (\ref{Zn-det-3}) and
(\ref{Zn-det-3d}).

The determinant (\ref{Zn-det-3d}) is already very close to the
desired form (\ref{Zn-det-4}), the only difference lying in the
upper index of the Laguerre functions ($L_{n+\ell-k}^{(k-\ell)}$
vs $L_{n+\ell-k}^{(k)}$). To demonstrate that this difference does
not affect the value of the determinant, we will constantly use
the identity
\begin{eqnarray}
    \label{id-4}
    L_{\nu}^{(\lambda-1)}(w) + L_{\nu-1}^{(\lambda)}(w) = L_{\nu}^{(\lambda)}(w)
\end{eqnarray}
when performing a set of row operations under the sign of
determinant in (\ref{Zn-det-3d}) as detailed below.

As the first row in (\ref{Zn-det-3d}) and the first row in
(\ref{Zn-det-4}) are already equivalent to each other, we start
with the last, the $N$-th row in (\ref{Zn-det-3d}). Adding to it
the $(N-1)$-th row raises the upper index of the $N$-th row
entries by $1$. In the next step, we add the content of the
$(N-2)$-th row to the $(N-1)$-th row; this also raises the upper
index of the entries in the $(N-1)$-th row by $1$. We go on with
this procedure up unless we arrive at the row $\sharp \; 2$ to
which we add the content of the row $\sharp \; 1$. This procedure
(consisting of $N-1$ steps) has `repaired' the row $\sharp \; 2$
bringing it to the form of the row $\sharp \; 2$ appearing in
(\ref{Zn-det-4}).

To proceed further, we come down to the last, $N$-th, row in
(\ref{Zn-det-3d}) adding to it the content of the previous,
$(N-1)$-th row. Again, we go on with this procedure up until we
reach the row $\sharp \; 4$ to which we add the content of the row
$\sharp \; 3$. This makes the row $\sharp \; 3$ fit the same row
in (\ref{Zn-det-4}). This took $(N-2)$ steps.

Now we again come down to the very last row and repeat the same
procedure. In $(N-3)$ steps we will find that it brings the row
$\sharp \; 4$ to the desired form. Proceeding in the same manner,
we realise that $N(N-1)/2$ row operations which do not change the
value of the determinant (\ref{Zn-det-3d}) bring the matrix under
the sign of determinant in (\ref{Zn-det-3d}) to that in
(\ref{Zn-det-4}). This completes the proof of the formula
(\ref{Z-equiv}).


\begin{references}


\bibitem{EA-1975}
    S. F. Edwards and P. W. Anderson,
    Theory of spin glasses,
    J. Phys. F: Met. Phys. {\bf 5}, 965 (1975).

\bibitem{Remark-example} See (\ref{Zn-mapped}) below for immediate example of
    a fermionic replica filed theory formulated in terms of some matrix field
    ${\cal Q} \in {\mathbb C}^{n \times n}$. Explanations will
    follow.

\bibitem{W-1979}
    F. Wegner,
    The mobility edge problem: Continuous symmetry and a conjecture,
    Z. Phys. B {\bf 35}, 207 (1979).

\bibitem{SW-1980}
    L. Sch\"afer and F. Wegner,
    Disordered system with $n$ orbitals per site: Lagrange
    formulation, hyperbolic symmetry, and Goldstone modes,
    Z. Phys. B {\bf 38}, 113 (1980).

\bibitem{ELK-1980}
    K. B. Efetov, A. I. Larkin, and D. E. Khmelnitskii,
    Interaction between diffusion modes in localization theory,
    Zh. \'Eksp. Teor. Fiz. {\bf 79}, 1120 (1980)
    [Sov. Phys. JETP {\bf 52}, 568 (1980)].

\bibitem{P-2002} G. Parisi,
    Two spaces looking for a geometer, Bull. Symbolic Logic {\bf 9},
    181 (2003).

\bibitem{K-2003} E. Kanzieper,
    Replica sigma models and nonlinear integrable hierarchies,
    unpublished (2003).

\bibitem{VZ-1985}
    J. J. M. Verbaarschot and M. R. Zirnbauer,
    Critique of the replica trick,
    J. Phys. A: Math. and Gen. {\bf 17}, 1093 (1985).

\bibitem{KM-1999a}
    A. Kamenev and M. M\'ezard,
    Wigner-Dyson statistics from the replica method,
    J. Phys. A: Math. and Gen. {\bf 32}, 4373 (1999).

\bibitem{KM-1999b}
    A. Kamenev and M. M\'ezard,
    Level correlations in disordered metals: the replica
    $\sigma$-model,
    Phys. Rev. B {\bf 60}, 3944 (1999).

\bibitem{YL-1999}
    I. V. Yurkevich and I. V. Lerner,
    Nonperturbative results for level correlations from the
    replica nonlinear $\sigma$ model,
    Phys. Rev. B {\bf 60}, 3955 (1999).

\bibitem{Z-1999} M. R. Zirnbauer,
    Another critique of the replica trick,
    e-print cond-mat/9903338 (1999).

\bibitem{K-2001} E. Kanzieper, Random matrices and the replica
    method, Nucl. Phys. B {\bf 596}, 548 (2001).

\bibitem{Remark-perturbative} In the perturbative region of the field theory, the replica
    parameter $n$ merely serves as a book-keeping tool identifying
    unphysical vacuum loops in a diagrammatic expansion for the
    partition function $Z_n$. Due to a finite number of expansion
    terms being taken into account, the dependence of $Z_n$ on $n$
    is algebraic or rational at most making the analytic continuation away
    from $|n| \in {\mathbb Z^+}$ be straightforward.

\bibitem{Remark-Keldysh} Recently formulated Keldysh field theory
    is a possible alternative. See: A. Kamenev and A. Andreev,
    Electron-electron interactions in disordered metals: Keldysh
    formalism,
    Phys. Rev. B {\bf 60}, 3944 (1999). Also, see the latest development due to
    Schwiete and Efetov [e-print cond-mat/0409546] who showed how a SUSY field
    theory can be extended to account for the interplay between disorder and
    interaction to the {\it first} order in electron-electron interaction and
    to all orders in disorder.

\bibitem{M-1991}
    M. L. Mehta,
    {\it Random Matrices}
    (Academic Press, New York, 1991).

\bibitem{E-book} K. Efetov, {\it Supersymmetry in Disorder and Chaos}
    (CUP, Cambridge, 1997).

\bibitem{MPV-1987}
    M. M\'ezard, G. Parisi, and M. A. Virasoro,
    {\it Spin Glass Theory and Beyond}
    (World Scientific, Singapore, 1987).

\bibitem{GK-2001} D. M. Gangardt and A. Kamenev,
    Replica treatment of the Calogero-Sutherland model,
    Nucl. Phys. B {\bf 610}, 578 (2001).

\bibitem{NGK-2002} S. M. Nishigaki, D. M. Gangardt, and
    A. Kamenev,
    Correlation functions of the BC Calogero-Sutherland model,
    J. Phys. A: Math. and Gen. {\bf 36}, 3137 (2003).

\bibitem{DV-2001} D. Dalmazi and J. J. M. Verbaarschot,
    The replica limit of unitary matrix integrals,
    Nucl. Phys. B {\bf 592}, 419 (2001).

\bibitem{NK-2002} S. M. Nishigaki and A. Kamenev,
    Replica treatment of non-Hermitian disordered Hamiltonians,
    J. Phys. A: Math. and Gen. {\bf 35}, 4571 (2002).

\bibitem{K-2002} E. Kanzieper,
    Replica field theories, Painlev\'e transcendents and
    exact correlation functions,
    Phys. Rev. Lett. {\bf 89}, 250201 (2002).

\bibitem{T-2000} G. Teschl,
    {\it Jacobi Operators and Completely Integrable Nonlinear
    Lattices}
    (AMS, Providence, RI, 2000).

\bibitem{O-1} K. Okamoto,
    Studies on the Painlev\'e equations, I. Sixth Painlev\'e equation PVI,
    Ann. Mat. Pura Appl. (4), {\bf 146}, 337 (1987).

\bibitem{O-2} K. Okamoto,
    Studies on the Painlev\'e equations, II. Fifth Painlev\'e
    equation PV,
    Japan. J. Math. {\bf 13}, 47 (1987).

\bibitem{O-3} K. Okamoto,
    Studies on the Painlev\'e equations, III. Second and fourth
    Painlev\'e equations, PII and PIV,
    Math. Ann. {\bf 275}, 221 (1986).

\bibitem{O-4} K. Okamoto,
    Studies on the Painlev\'e equations, IV. Third Painlev\'e
    equation PIII,
    Funkcial. Ekvac. {\bf 30}, 305 (1987).

\bibitem{SV-2003} K. Splittorff and J.J.M. Verbaarschot,
    Replica limit of the Toda lattice equation,
    Phys. Rev. Lett. {\bf 90}, 041601 (2003).

\bibitem{SV-2003new} K. Splittorff and J.J.M. Verbaarschot,
    Factorisation of correlation functions and the replica limit
    of the Toda lattice equation, Nucl. Phys. B {\bf 683}, 467 (2004).

\bibitem{New-Remark} In particular, {\it both} compact (fermionic) and
    non-compact (bosonic) integrals for partition functions appear
    explicitly in the formalism of Ref. \cite{SV-2003new}.

\bibitem{G-1965} J. Ginibre,
    Statistical ensembles of complex quaternion and real matrices,
    J. Math. Phys. {\bf 6}, 440 (1965).

\bibitem{MWZ-2000} M. Mineev-Weinstein, P. B. Wiegmann, and A.
    Zabrodin,
    Integrable structure of interface dynamics,
    Phys. Rev. Lett. {\bf 84}, 5106 (2000).

\bibitem{ABWZ-2002} O. Agam, E. Bettelheim, P. Wiegmann, and A.
    Zabrodin,
    Viscous fingering and a shape of an electronic droplet in the Quantum Hall regime,
    Phys. Rev. Lett. {\bf 88}, 236801 (2002).

\bibitem{Z-2002} A. Zabrodin,
    New applications of non-Hermitean random matrices, Annales Henri Poincare {\bf 4}, S851,
    Suppl. 2 (2003).

\bibitem{FS-2003} Y. V. Fyodorov and H.-J. Sommers,
    Random matrices close to Hermitean or unitary: Overview of methods and results,
    J. Phys. A: Math. and Gen. {\bf 36}, 3303 (2003).

\bibitem{Remark-NC} The normalisation constant is fixed by the
    integration measure $\prod_{k,\,\ell=1}^N d{\rm Re}\, H_{k\ell} \;
    d{\rm Im}\, H_{k\ell}$.

\bibitem{Remark-Delta} The $\delta$--function in the complex plane
    is understood as $\delta^2(z) = \delta ({\rm Re\,}z)\, \delta ({\rm
    Im\,}z)$.

\bibitem{Remark-Zn} In the regime in question, the parameter $N$ enters
    ${\tilde Z}_n$ [see (\ref{dd-nfold})] as a prefactor vanishing in the
    replica limit. For this reason, we will sometimes drop $N$ from the
    arguments of ${\tilde Z}_n$. This is also the reason why $N$ does not
    appear as a parameter in the l.h.s. of (\ref{dd-replica-zw}).

\bibitem{Remark-Parisi} Unless we know how to interprete\cite{P-2002} the
    integration measure at $n \notin {\mathbb Z}^+$.

\bibitem{A-1883} C. Andr\'eief,
    Note sur une relation les int\'egrales d\'efinies des produits des fonctions,
    M\'em. de la Soc. Sci., Bordeaux {\bf 2}, 1 (1883).

\bibitem{dB-1955} N. G. de Bruijn,
    On some multiple integrals involving determinants,
    J. Indian Math. Soc. {\bf 19}, 133 (1955).

\bibitem{D-1972} G. Darboux,
    {\it Lecons sur la Theorie Generale des Surfaces et les Applications
    Geometriques du Calcul Infinitesimal}, Vol. II: XIX  (Chelsea, New York, 1972).

\bibitem{F-1983}
    A. M. Finkelstein,
    Influence of Coulomb interaction on the properties of disordered metals,
    Zh. \'Eksp. Teor. Fiz. {\bf 84}, 168 (1983)
    [Sov. Phys. JETP {\bf 57}, 97 (1983)].

\bibitem{FW-2002} P. J. Forrester and N. S. Witte,
    Application of the $\tau$-function theory of Painlev\'e equations to
    random matrices: PV, PIII, the LUE, JUE, and CUE,
    Commun. Pure Appl. Math. {\bf 55}, 679 (2002).

\bibitem{TW-1994a} C. A. Tracy and H. Widom,
    Fredholm determinants,  differential equations and matrix
    models,
    Commun. Math. Phys. {\bf 163}, 33 (1994).

\bibitem{TW-1994b} C. A. Tracy and H. Widom,
    Level spacing distributions and the Bessel kernel,
    Commun. Math. Phys. {\bf 161}, 289 (1994).

\bibitem{JMMS-1980} M. Jimbo, T. Miwa,Y. M\^ori, and M. Sato,
    Density matrix of an impenetrable Bose gas and the fifth Painlev\'e
    transcendent,
    Physica D {\bf 1}, 80 (1980).

\bibitem{F-book} P. Forrester, {\it Log-Gases and Random
    Matrices}, Ch. 6 (book in progress, 2002).

\bibitem{Remark-BCs} In fact, we do not actually need the boundary
    conditions [(\ref{PV-bc}) or (\ref{PIV-bc})]: general arguments based on existence, convergence
    and conservation laws do the job.

\bibitem{Remark-BCs-check} One may verify that
    $\sigma_{\rm V}(t)=n\, f_1(t;N)+ \cdots$ obeys the boundary condition
    (\ref{PV-bc}) up to the terms linear in $n$:
    \begin{eqnarray}
    \left. \sigma_{\rm V}(t)\right|_{t\rightarrow +\infty} \sim
    - nt + nN. \nonumber
    \end{eqnarray}

\bibitem{FW-2001} P. J. Forrester and N. S. Witte,
    Application of the $\tau$-function theory of Painlev\'e equations to
    random matrices: PIV, PII and the GUE,
    Commun. Math. Phys. {\bf 219}, 357 (2001).

\bibitem{Remark-Proof} This assumption might have been proven by
    examining (\ref{PV-eq}) in the large--$N$ scaling limit specified
    prior to (\ref{Zn-dos-tails}).

\bibitem{Remark-Expan} A term with $p=0$ is forbidden by the normalisation
    ${\tilde Z}_0^{\rm (tails)}=1$.

\bibitem{Remark-BCs-g1} Indeed, as the density of states $R_1^{\rm (tails)}(u)$ at the
    edge equals $R_1^{\rm (tails)}(u) = R_1(z=(\sqrt{N}+u)e^{i\varphi};N)$,
    the conservation constraint $\int_{\mathbb C} d^{\,2}z \, R_1(z;N) = N$
    translates to
    \begin{eqnarray}
        \label{cc-tails}
        \int_{-1}^\infty d\tau \, (1+\tau) \, R_1^{\rm (tails)}(\tau
        \sqrt{N})= \frac{1}{2\pi}. \nonumber
    \end{eqnarray}
    Taking the $N \rightarrow \infty$ limit and appealing to (\ref{R1-g1})
    yields the statement (\ref{g1-prime-prop}).

\bibitem{Remark-BCs-tails} Given $C_1=-7/4$ and $C_2=1/4$, one may verify
    that $\sigma_{\rm IV}(t)=n\, g_1(t;0)+ \cdots$ with $g_1(t)$
    in the form (\ref{g1-sol-plus}) obeys the boundary condition (\ref{PIV-bc})
    up to the term linear in $n$:
    \begin{eqnarray}
        \left. \sigma_{\rm IV}(t) \right|_{t \rightarrow -\infty}
        \sim -2nt.
    \end{eqnarray}

\bibitem{B-2002} J. Baik,
    Painlev\'e expressions for LOE, LSE and interpolating
    ensembles,
    Int. Math. Res. Not. {\bf 2002}, 1739 (2002).

\bibitem{Remark-BCs-R} One may verify that so derived
    $\sigma_{\rm V}(t)=n^2 h_2(t) + \cdots$ meets the
    small--$n$ version of the boundary condition (\ref{PV-bc2}):
    \begin{eqnarray}
    \left. \sigma_{\rm V}(t)\right|_{t\rightarrow \infty} \sim
    \frac{n^2}{t}\, e^{-t}. \nonumber
    \end{eqnarray}

\bibitem{Remark-ZJ} This representation can be proven along the lines of
    Ref. \cite{ZJ-1998} where a similar matrix representation appears
    for a scalar kernel in the context of Hermitean random matrices.

\bibitem{ZJ-1998} P. Zinn-Justin,
    Universality of correlation functions of Hermitean random
    matrices in an external field,
    Commun. Math. Phys. {\bf 194}, 631 (1998).

\bibitem{Remark-GUE} Similarly, for the matrix Hamiltonian ${\cal H} \in {\rm
    GUE}_N$ taken from the Gaussian Unitary Ensemble, one easily derives the
    following relation
    \begin{eqnarray}
        R_1(\epsilon;N) \propto e^{-\epsilon^2} \, {\tilde Z}_2(\epsilon;N-1)
        \nonumber
    \end{eqnarray}
    between the density of states $R_1(\epsilon;N)$ in the ${\rm
    GUE}_N$ and the fermionic replica partition function ${\tilde
    Z}_2(\epsilon;N-1)$ for the ${\rm GUE}_{N-1}$.

\bibitem{CDZ-2003} G. Carlet, B. Dubrovin, and Y. Zhang,
    The extended Toda hierarchy, Moscow Math. Journ. {\bf 4}, 313 (2004).

\bibitem{Jac-Comment} I thank Jac Verbaarschot for pointing out to
    me that evaluation of noncompact integrals \cite{New-Remark} can
    be traded for solving an appropriate Painlev\'e equation
    \cite{K-2002} taken at negative values of the replica index
    $n$. See, also, a remark below (21) in Ref. \cite{SV-2003}.

\bibitem{KAA-2000} I. L. Kurland, I. L. Aleiner, and B. L. Altshuler,
    Mesoscopic magnetization fluctuations for metallic grains close to
    the Stoner instability,
    Phys. Rev. B {\bf 62}, 14886 (2000).

\bibitem{OBWH-2001} Y. Oreg, P. W. Brouwer, X. Waintal, and B. I. Halperin,
    Spin, spin-orbit, and electron-electron interactions in mesoscopic
    systems, in: {\it Nano-Physics and Bio-Electronics: a New Odyssey},
    edited by T. Chakraborty, F. Peeters, and U. Sivan (Elsevier,
    2002).

\bibitem{ABG-2002} I. L. Aleiner, P. W. Brouwer, and L. I. Glazman,
    Quantum effects in Coulomb blockade,
    Phys. Rep. {\bf 358}, 309 (2002).

\end{references}
\end{document}